\documentclass[useAMS,usenatbib]{mn2e}   
\usepackage{times} 
\usepackage{psfig,epsfig} 
\usepackage{natbib} 
 
\def\aap{A\&A} 
\def\apj{ApJ} 
 
\def\apjl{ApJL} 
\def\mnras{MNRAS} 
\def\araa{ARA\&A}

\def\apjs{ApJS} 
\def\gca{Geochimica et Cosmochimica Acta}

\newcommand{\hm}{\,h^{-1}{\rm Mpc}}  
\newcommand{\chandra}{{\it Chandra} }

\newcommand{\xmas}{{\it X-MAS} } 
\newcommand{\be}{\begin{equation}} 
\newcommand{\ee}{\end{equation}} 
\newcommand{\ba}{\begin{eqnarray}} 
\newcommand{\ea}{\end{eqnarray}} 
\newcommand{\nn}{\nonumber} 
   
\title[Systematics in the X-ray Cluster Mass Estimators]  
{Systematics in the X-ray Cluster Mass Estimators}  
  \author[E. Rasia et al.]   
{E. Rasia$^1$, S. Ettori$^2$,  L. Moscardini$^{3}$, P. Mazzotta$^{4,5}$,  
S. Borgani$^{6,7,8}$, K. Dolag$^{9}$,    \\~\\ 
\LARGE{\rm  G. Tormen$^1$, L.M. Cheng$^{10}$, A. Diaferio$^{11}$} \\~\\   
\footnotesize   
$^1$ Dipartimento di Astronomia, Universit\`a di Padova, vicolo  
  dell'Osservatorio 2, I-35122 Padova, Italy (rasia,tormen@pd.astro.it)\\  
$^2$ INAF, Osservatorio Astronomico di Bologna, via Ranzani 1, I-40127  
Bologna, Italy (stefano.ettori@bo.astro.it)\\  
$^3$ Dipartimento di Astronomia, Universit\`a di Bologna, 
via Ranzani 1, I-40127 Bologna, Italy 
(lauro.moscardini@unibo.it)\\ 
$^4$ Dipartimento di Fisica, Universit\`a di Roma Tor Vergata, 
via della Ricerca Scientifica 1, I-00133 Roma, Italy 
(pasquale.mazzotta@roma2.infn.it)\\ 
$^{5}$ Harvard-Smithsonian Centre for Astrophysics, 60 Garden Street, 
Cambridge, MA 02138, USA\\ 
$^6$ Dipartimento di Astronomia, Universit\`a di Trieste, via 
  Tiepolo 11, I-34131 Trieste, Italy (borgani@ts.astro.it)\\ 
$^7$ INFN -- National Institute for Nuclear Physics, Trieste, 
  Italy\\
$^8$ INAF --  Osservatorio Astronomico di Trieste, via Tiepolo 11, I-34131,
 Trieste, Italy\\
$^{9}$ Max-Planck-Institut f\"ur Astrophysik, Karl-Schwarzschild Strasse 
  1, D-85748 Garching bei M\"unchen, Germany (kdolag@mpa-garching.mpg.de) \\
$^{10}$ Institute of Theoretical Physics, Chinese Accademy of Sciences, Beijing 
  100080, China (clm@itp.ac.cn)\\ 
$^{11}$ Dipartimento di Fisica Generale ``Amedeo Avogadro'', Universit\`a di 
  Torino, via Giuria 1, I-10125, Torino, Italy (diaferio@ph.unito.it)
}  
 
\date{1st draft: jul 05}  
\begin{document}   
\maketitle   
  
\begin{abstract}   
 
We examine the systematics affecting the X-ray mass estimators applied to a
set of five galaxy clusters resolved at high resolution in hydrodynamic
simulations, including cooling, star formation and feedback processes.  These
simulated objects are processed through the X-ray Map Simulator, \xmas,
to provide \chandra-like long exposures that are analyzed to reconstruct
the gas temperature, density, and mass profiles used as input. These clusters
have different dynamic state: we consider an hot cluster with temperature
$T=11.4$ keV, a perturbed cluster with $T=3.9$ keV, a merging object with
$T=3.6$ keV, and two relaxed systems with $T=3.3$ keV and $T=2.7$ keV,
respectively. These systems are located at $z=0.175$ so that their emission
fits within the \chandra ACIS-S3 chip between 0.6 and 1.2 $R_{500}$.  

We find that the mass profile obtained via a direct application of the
hydrostatic equilibrium equation is dependent upon the measured temperature
profile.  An irregular radial distribution of the temperature values, with
associated large errors, induces a significant scatter on the reconstructed
mass measurements.  At $R_{2500}$, the actual mass is recovered within 1
$\sigma$, although we notice this estimator shows high statistical errors due
to high level of \chandra background.  Instead, the poorness of the
$\beta-$model in describing the gas density profile makes the evaluated masses
to be underestimated by $\sim 40$ per cent with respect to the true mass, both
with an isothermal and a polytropic temperature profile.  We also test ways to
recover the mass by adopting an analytic mass model, such as those proposed by
\cite{1997ApJ...490..493N} and \cite{2004MNRAS.351..237R}, and fitting the
temperature profile expected from the hydrostatic equilibrium equation to the
observed one.  We conclude that the methods of the hydrostatic equilibrium
equation and those of the analytic fits provide a more robust mass estimation
than the ones based on the $\beta-$model.  In the present work the main
limitation for a precise mass reconstruction is to ascribe to the relatively
high level of the background chosen to reproduce the \chandra one.  After
artificially reducing the total background by a factor of 100, we find that
the estimated mass significantly underestimates the true mass profiles. This
is manly due (i) to the neglected contribution of the gas bulk motions to the
total energy budget and (ii) to the bias towards lower values of the X-ray
temperature measurements because of the complex thermal structure of the
emitting plasma.
\end{abstract}   
   
 
\section{INTRODUCTION}   
  
The gravitating mass in galaxy clusters is the most fundamental 
quantity to use them as cosmological probes.  Measurements of the 
total cluster masses have been used for the last 70 years  
\citep[since][]{zwicky1933} to infer the presence of dark matter over Mpc 
scales and to constrain the cosmological parameters that describe the 
distribution and growth of the cosmic density fluctuations of 
$>10^{14} M_{\odot}$ virialized structures 
\citep[see the reviews by][]{rosati2002,voit2005}.  These structures appear 
as well defined and easily observable at the X-ray wavelengths, where the 
emission is mainly dominated by bremsstrahlung processes and is 
proportional to the square of the plasma density. This observational 
evidence allows to identify massive relaxed clusters and to study 
their physical properties through the estimates of the X-ray emitting 
intra-cluster medium (ICM) density and temperature.  Since the sound 
crossing time in the ICM is sufficiently shorter than the age of the 
structure, the assumption that the ICM is in hydrostatic equilibrium 
within the underlying dark matter potential generally holds 
\citep{2002mpgc.book....1S}.  ICM density and temperature are  used 
to determine the radial mass profile via the hydrostatic equation 
\citep[see, e.g.,][]{1987ApJ...317..593C}, with a further  
assumption of the spherical symmetry for the sake of simplicity. 
  
In the present work, we investigate the systematics that affect the X-ray mass
measurements by using hydrodynamic simulations processed by our X-ray Map
Simulator \citep[\xmas,][]{2004MNRAS.351..505G} to produce realistic
event files and images that are then analyzed in a way identical to what is
done with observed clusters systems. Several authors studied the robustness of
the hydrostatic equilibrium and the uncertainties related to this mass
estimator.  General conclusions were that there is a good agreement between
X-ray and true masses, in particular in the outer regions, with a standard
deviation of approximately 15 per cent (see, e.g.,
\citealt{1996A&A...305..756S,1996ApJ...469..494E}; cfr. also
\citealt{bartelmann1996}).  \citet{1997ApJ...487...33B} pointed out, however,
that the uncertainties on the mass measurement propagated from the weak
constraints on the gas temperature profile via the hydrostatic equilibrium
equation make any accuracy on the determination of the mass profile very poor.
More recently, \citet{2004MNRAS.351..237R} \citep[see
also][]{2004MNRAS.355.1091K,borgani2004} found in a set of hydro-N-body
simulations that the ICM is not in a perfect hydrostatic state and, thus, the
masses can be underestimated by up to 20 per cent in non-radiative ICM models
due to residual gas bulk motions. \citet{2004MNRAS.355.1091K} conclude that
more thermalization (and less significant departures from the true mass
estimates) is obtained when cooling and feedback are included and that an
isothermal $\beta$--model still provides accurate mass estimates within 20-30
per cent uncertainties.  With our approach in which (1) the known input on the
matter distribution is properly convolved with the response of real
instruments and (2) an observational-like analysis is performed on the output
to recover the original input, we extend the previous work limited to the
study of the hydrodynamic simulations in order to highlight the systematic
effects present both in the assumption of hydrostatic equilibrium and in the
observational techniques adopted in recovering the mass estimates.
 
The paper is organized as follows. We introduce our simulated dataset 
in Section~2.  The outputs produced by combing the hydrodynamic 
simulations and the X-ray Map Simulator for \chandra data are 
presented in Section~3 and analyzed to recover the gravitating mass in 
Section~4.  The results are presented in Section~5 and summarized and 
discussed in Section~6.  All the errors quoted are at $1 \sigma$ level 
(68.3 per cent level of confidence for one interesting parameter). 
 
\section{THE SAMPLE OF SIMULATED CLUSTERS} 
\label{sec:simul}

The simulated clusters have been taken from two different simulations in order
to obtain a sample with a significant range of masses and dynamical states.

The simulations has been carried out 
with {\small GADGET-2} \citep{springel2005}, a new version of the 
parallel Tree-SPH simulation code {\small GADGET} \citep{SP01.1}.  It 
uses an entropy-conserving formulation of SPH 
\citep{2002MNRAS.333..649S}, and includes radiative cooling, heating 
by a UV background, and a treatment of star formation and feedback 
processes. The latter is based on a sub-resolution model for the 
multi-phase structure of the interstellar medium \citep{springel2003}.

The first cluster, $C_{Hot}$, has been extracted from a dark-matter only
simulation with a box-size of 479 $h^{-1}$ Mpc \citep{yoshida.etal.01}.  The
assumed cosmological model is a standard flat $\Lambda$CDM universe, with
$\Omega_{\rm m} = 1-\Omega_{\Lambda} = 0.3$, $\sigma_8 = 0.9$, $\Omega_{\rm b}
h^2 = 0.019$ and $H_0 = 100 h$ km s$^{-1}$ Mpc$^{-1}$ with $h=0.7$. This
cluster has been resimulated at higher mass and force resolution. The new
initial conditions for this system have been generated by applying the Zoomed
Initial Condition (ZIC) technique \citep{1997MNRAS.286..865T}. This method
allows one to increase the mass resolution in a suitably chosen
high--resolution Lagrangian region surrounding the structure to be
re--simulated, and at the same time to correctly describe the large--scale
tidal field of the cosmological environment by using low--resolution
particles. The
mass resolution of the gas particle for this cluster is $1.7 \times 10^8
h^{-1}M_{\odot}$, the gravitational softening length was kept fixed at
$\epsilon=30.0 h^{-1}$kpc comoving (Plummer-equivalent) and was switched to a
physical softening length of $\epsilon= 5.0 h^{-1}$ kpc at $z=5$. The
resimulation follows star formation, feedback and heating by thermal
conduction \citep{2004MNRAS.351..423J,dolag.etal.04}. The SN efficiency in
powering galactic winds is set to 50 per cent which turns into a wind speed of
340 km s$^{-1}$.

The other four simulated systems, $C_{Pert}$, $C_{Merg}$, $C_{Rel1}$,
and $C_{Rel2}$, used in the following analysis, have been extracted from the
large--scale hydrodynamic simulation described in \cite{borgani2004}. We refer
to that paper for a detailed description of that simulation, while we provide
here only a short summary. 
 The simulation follows the evolution of 
$480^3$ dark matter (DM) particles and an initially equal number of 
gas particles within a box of $192\hm$ on a side of the previous cosmological
model, but with a lower normalization of the power spectrum, $\sigma_8=0.8$.
The mass of the DM and gas particles are, respectively, $m_{\rm 
DM}=4.6\times 10^9 h^{-1} M_\odot$ and $m_{\rm gas}=6.9\times 10^8 
h^{-1} M_\odot$. The Plummer--equivalent gravitational softening was kept
fixed at $\epsilon_{\rm Pl}=7.5\,h^{-1}$ kpc comoving and  
switched in physical units at $z=2$.  
 
$C_{Pert}$ is directly extracted from the final output of this simulation by
using the standard identification criterion based on the spherical
over-density.  From the same simulation we selected the remaining three objects
which, in an analogous way of $C_{Hot}$, have been re-simulating at
high-resolution, following the ZIC tecnique.
 
These runs, whose results are extensively discussed elsewhere
\citep{borgani.new}, have been performed by assuming a mass for the DM and gas
particles 10 times smaller than in the original simulation; also the
Plummer--equivalent gravitational softening has been reduced and corresponds
to $\epsilon_{\rm Pl}=3.5\,h^{-1}$ kpc at $z=0$.  For these new runs the
feedback scheme is calibrated to lead to higher galactic wind velocities with
respect to the original \cite{borgani2004}' simulation ($\approx 480$ km/s
instead of $\approx 360$ km/s).
 
The main properties of the five clusters forming our dataset are 
summarized in Table~\ref{tab:cl}. In particular we list: the virial 
radius, $R_{\rm vir}$, defined as the radius at which the over-density 
assumes the value dictated by the spherical top-hat model 
\citep[see, e.g., ][]{eke1996};  the radii $R_{2500}$ and $R_{500}$,  
within which the over-density with respect to the critical value is 
2500 and 500, respectively; the virial mass $M_{\rm vir}$, i.e. the 
mass included inside $R_{\rm vir}$; the number of DM and gas particles 
inside the virial radius ($N_{\rm DM}$ and $N_{\rm gas}$, 
respectively); the spectroscopic-like temperature, defined as 
$T_{\rm sl} \equiv \int W T dV/\int W dV$, where the weighting 
function $W$ equals to $n^2 /T^{0.75}$, being $n$ and $T$ the density 
and temperature of each gas particle.  This temperature definition has 
been introduced in the analysis of the hydrodynamic simulations of 
galaxy clusters by \cite{2004MNRAS.354...10M} to provide a better 
approximation to the values extracted from fits of observed X-ray 
spectra \citep[see also][ for an extension to cooler systems considering the 
  effects of metals]{vikhlinin2005}. 
 
To summarize, the following five objects have been selected as 
examples of clusters with different thermal and dynamic states: 
 
\begin{itemize} 
 
\item $C_{Hot}$ ($T_{\rm sl}=11.4$ keV, $M_{\rm vir}=2.26\times 10^{15}\, 
M_\odot$): a forming cluster, still accreting mass from the outskirts;

\item $C_{Pert}$ ($T_{\rm sl}=3.9$ keV, $M_{\rm vir}=7.0\times 10^{14}\, 
M_\odot$): a perturbed cluster which shows in the temperature map a 
cold substructure infalling toward the centre; 
 
\item $C_{Merg}$ ($T_{\rm sl}=3.6$ keV,  
$M_{\rm vir}=4.1\times 10^{14}\, M_\odot$): an object that experienced 
a recent major merger; 
 
\item $C_{Rel1}$ ($T_{\rm sl}=3.3$ keV, 
$M_{\rm vir}=3.6\times 10^{14}\, M_\odot$): a relaxed structure; 
 
\item $C_{Rel2}$ ($T_{\rm sl}=2.7$ keV, 
$M_{\rm vir}=2.3\times 10^{14}\, M_\odot$): a colder relaxed cluster.
 
\end{itemize}

\begin{table}  
\caption{ Properties of the simulated galaxy clusters forming our dataset
(their identification names are given in the corresponding columns). $R_{\rm
vir}$ is the virial radius; $R_{500}$ and $R_{2500}$ are the radii within
which the over-density with respect to the critical value is 500 and 2500,
respectively; $M_{\rm vir}$ is the virial mass. $N_{\rm DM}$ and $N_{\rm gas}$
represent the number of DM and gas particles inside $R_{\rm vir}$,
respectively. $T_{\rm sl}$ is the spectroscopic-like temperature inside
$R_{\rm vir}$, calculated by summing over the particles having $T>0.5$ keV
($T>1$ keV for $C_{Pert}$); $r_{\rm s,NFW}$ and $c_{\rm NFW}$ and $r_{\rm
s,RTM}$ and $c_{\rm RTM}$ are the scale radius (rescaled at the observation
redshift) and the concentration of the NFW and RTM mass model, respectively.}
 
\begin{tabular}{l c c c c c}  
\hline \\  
&$C_{Hot}$& $C_{Pert}$ & $C_{Merg}$ & $C_{Rel1}$ & $C_{Rel2}$ \\ 
$R_{\rm vir}\;[kpc]$ &2713&1967  & 1662 & 1567& 1368\\ 
$R_{500}\;[kpc] $    &1255&1225  & 808  & 732 & 646\\ 
$R_{2500}\;[kpc]$    &515 & 585  & 361  & 305 &265 \\ 
$M_{\rm vir}\;[10^{14} M_\odot]$ &22.6  & 7.0 & 4.1  & 3.6  & 2.3   \\
$N_{\rm DM}\;[10^5]$ &17.4& 0.9  & 5.5  & 4.6 & 3.0\\ 
$N_{\rm gas}\;[10^5]$&14.1& 0.7  & 4.2  & 3.9 & 2.6\\ 
$T_{\rm sl}\;[keV]$  &11.4& 3.9  & 3.6  & 3.3 & 2.7\\
&&&&&\\
$r_{\rm s,NFW}$ & 438  &250  & 182  & 168  & 142  \\
$c_{\rm NFW}  $ &  4.5 &5.0  & 5.3  & 5.4  & 5.3 \\
$r_{\rm s,RTM}$ & 170  &93   & 63   & 57   & 46 \\
$c_{\rm RTM}  $ & 11.7 &13.5 & 15.6 & 16.4 & 16.4 \\
\end{tabular}  
\label{tab:cl} 
\end{table}  
 
\begin{figure*} 
\psfig{figure=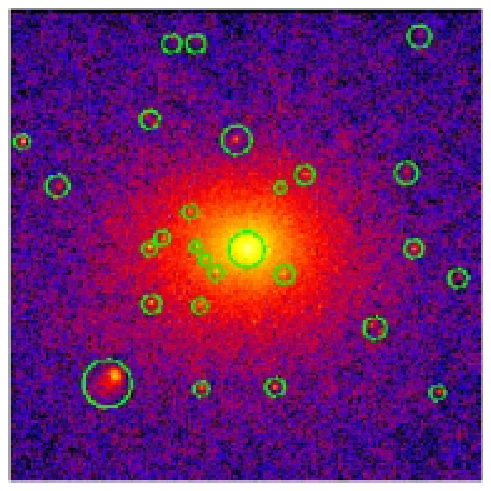,width=0.20\textwidth} 
\psfig{figure=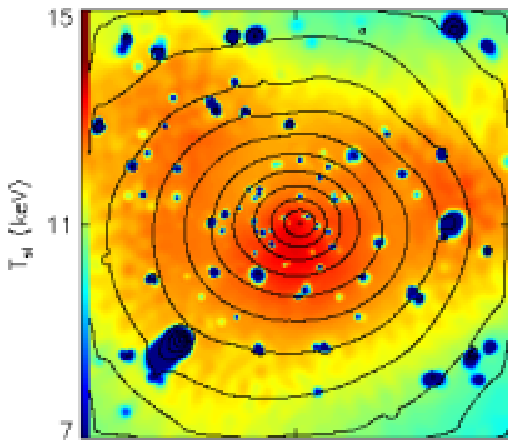,width=0.23\textwidth} 
\psfig{figure=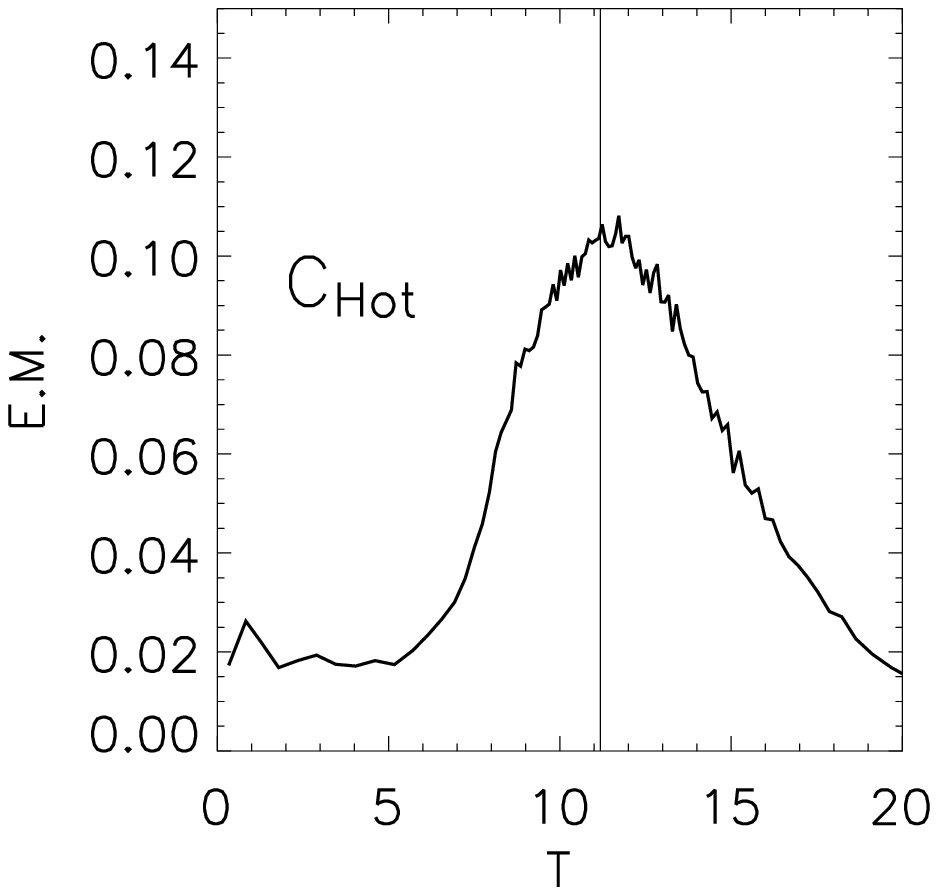,width=0.25\textwidth} 

\psfig{figure=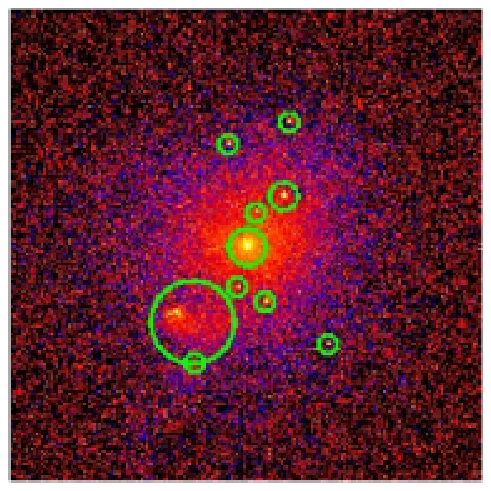,width=0.20\textwidth} 
\psfig{figure=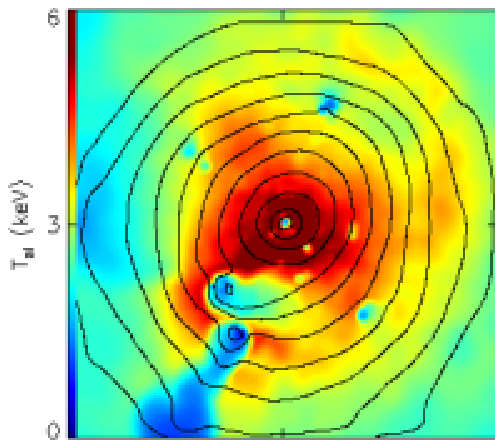,width=0.23\textwidth} 
\psfig{figure=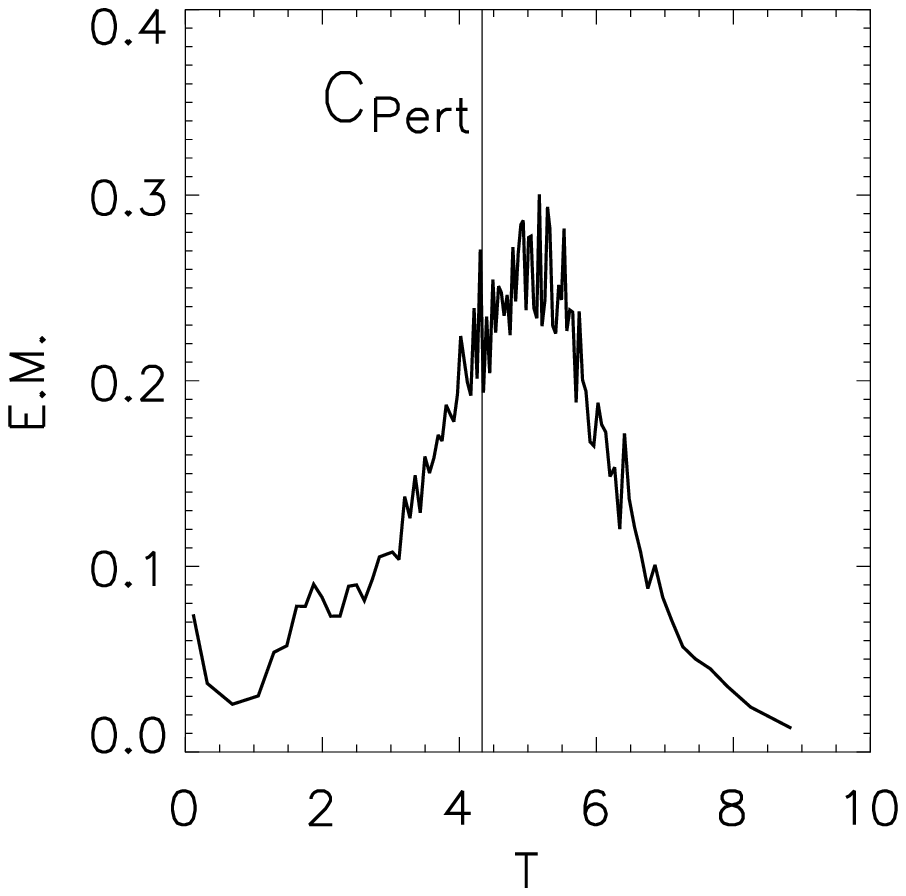,width=0.25\textwidth} 

\psfig{figure=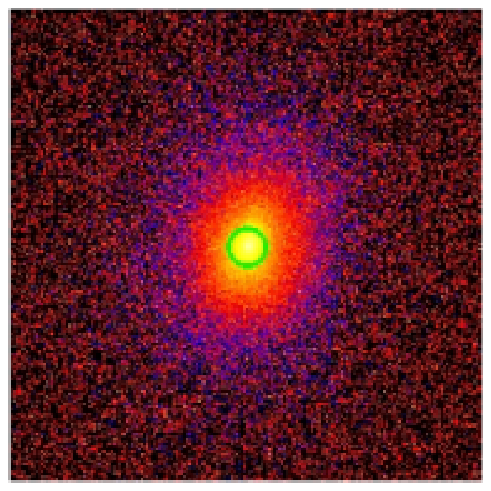,width=0.20\textwidth} 
\psfig{figure=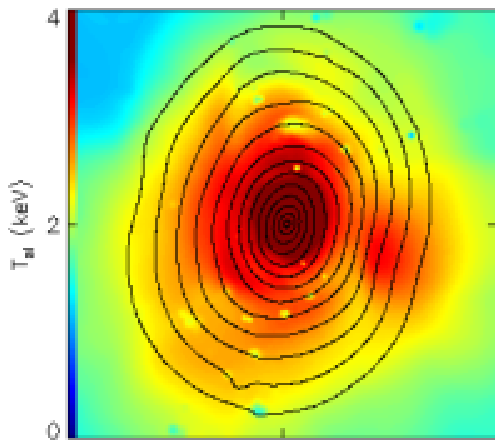,width=0.23\textwidth} 
\psfig{figure=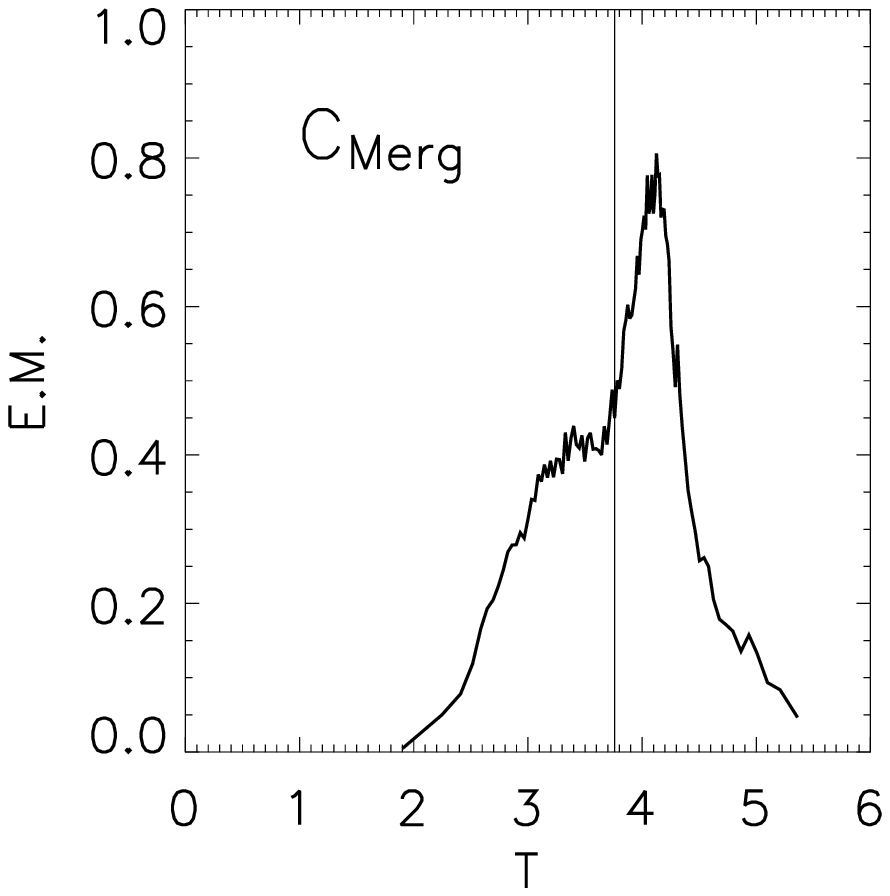,width=0.25\textwidth} 

\psfig{figure=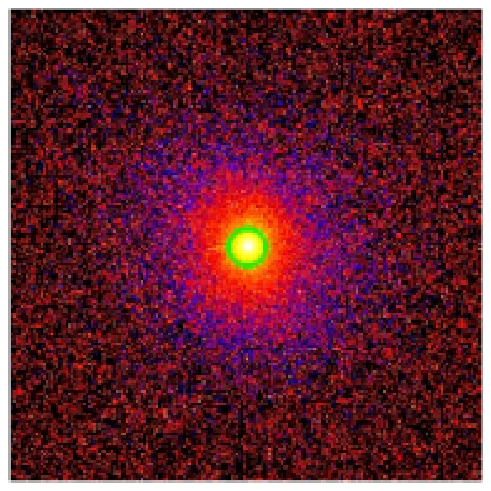,width=0.20\textwidth} 
\psfig{figure=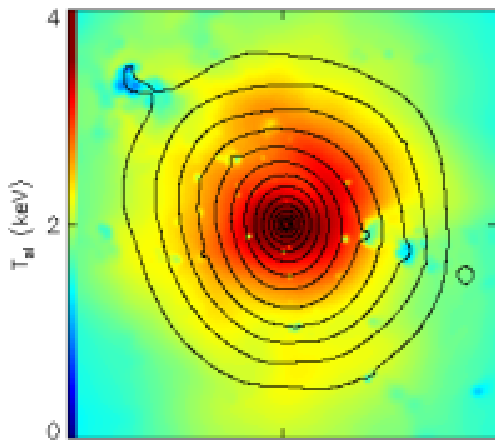,width=0.23\textwidth} 
\psfig{figure=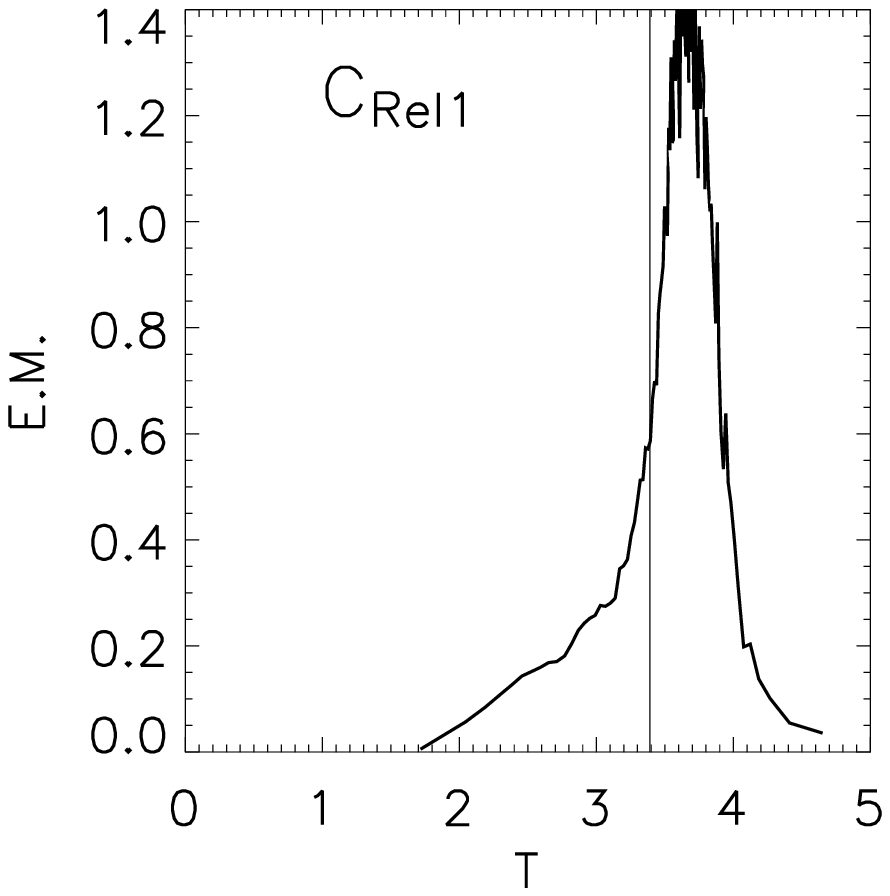,width=0.25\textwidth} 

\psfig{figure=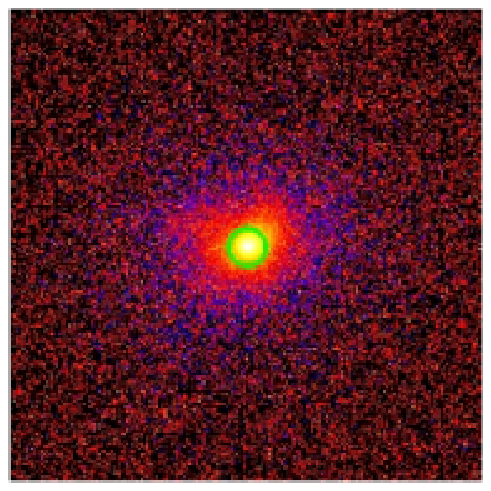,width=0.20\textwidth} 
\psfig{figure=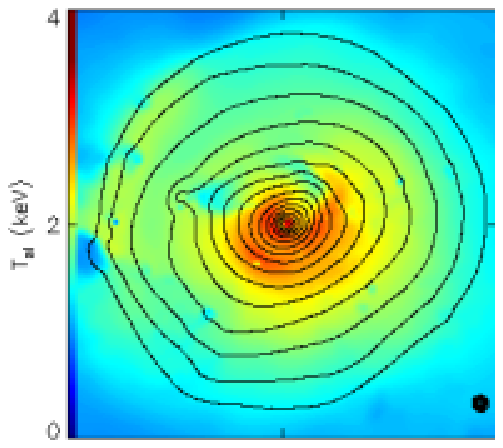,width=0.23\textwidth} 
\psfig{figure=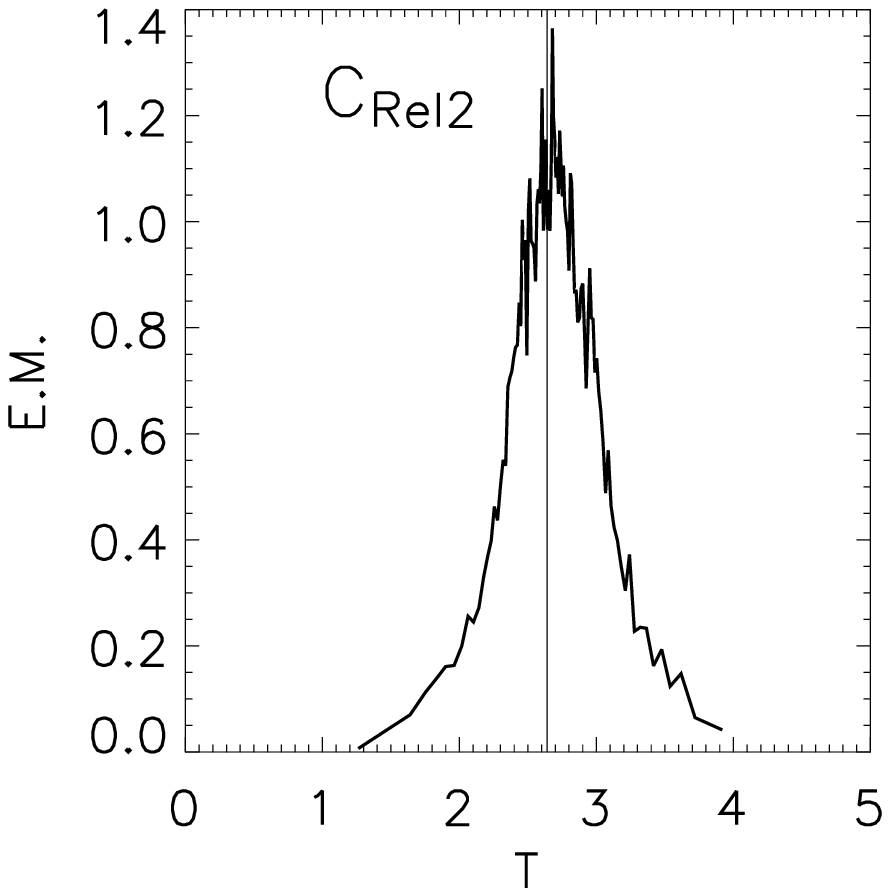,width=0.25\textwidth} 
 
\caption[]{ Images obtained from the simulated galaxy clusters: different rows
refer to $C_{Hot}$, $C_{Pert}$, $C_{Merg}$, $C_{Rel1}$ and $C_{Rel2}$, from
top to bottom.  Left column: photon images in the X-ray soft [0.5-2 keV]
energy band.  The images are 8.3 arcmin-width, exposure-corrected, and binned
to $2''$--pixel. The green circles show the regions masked out in the
analysis. Central column: X-ray logarithmic isoflux contours over-plotted to
the spectroscopic-like temperature map, both extracted directly from the
hydrodynamic simulations; the temperature scale is shown on the left.  Right
column: emission measure from the region inside $R_{500}$ (inside the box
photon image for $C_{Hot}$, $C_{Pert}$); the solid vertical black line refers
to $T_{2500}$ (see Table~\ref{tab:param}).}
\label{fig:photonimg1} 
\end{figure*} 
 
\section{Mock X-ray observations of the high-resolution simulated clusters} 
 
The sample of our hydro-N-body simulated galaxy clusters has been processed
with \xmas to obtain mock \chandra ACIS-S3 observations. The event files
produced with our program are then analyzed in the same way and with the same
tools of real observations. A detailed description of \xmas is reported in
\citet{2004MNRAS.351..505G}.  The design of the two main units constituting
the software is here summarized.
 
\begin{itemize} 
 
\item {\bf First Unit: generation of differential flux maps.}  The input of
this unit is the output of a hydro-N-body simulation.  For each gas particle,
we compute the emissivity and distribute it over the corresponding
volume. After selecting a line of sight for the simulated observation, we
compute the projected spectrum for each pixel and then the differential flux
for each angular coordinate in bins of energy. The final step is to add
Galactic absorption.
 
\item {\bf Second Unit: simulation of \chandra observations.}  We estimate the
 expected number counts and iteratively subdivide the tile region until the
 counts become smaller than a given threshold (10 counts); we then use the
 command \textsc{fakeit} in the utility XSPEC \citep[see, e.g., Xspec User's
 Guide version
 12.2\footnote{http://heasarc.gsfc.nasa.gov/docs/xanadu/xspec/XspecManual.pdf};][]{2003xspec}
 to convolve the spectral model of each subregion with the response of the CCD
 and to add the sky background (acis\_c\_S3\_bg\_evt\_191000.fits); at the end
 we generate the final photon event file.
 
\end{itemize} 
 
A Galactic equivalent column density of $N_H=5\times 10^{20}$ cm$^{-2}$ is
assumed and the cluster metallicity has been fixed at $Z=0.3\,Z_\odot$, using
the solar value tabulated in \cite{1989GeCoA..53..197A}.  In order to be
able to observe a physical size of $\sim$ 1.5~Mpc within the field of view
of \chandra ACIS-S3 (8.3 arcmin) we impose a redshift $z=0.175$ for the
simulated clusters.
 
Since we want to study the systematic discrepancies between observed and real
quantities for ideal observations, we have applied very long exposure times to
all the simulated observations aiming to minimize the statistical
uncertainties related to the number counts: 0.5 Msec for $C_{Hot}$, 1 Msec for
$C_{Pert}$, $C_{Merg}$, $C_{Rel1}$ and $C_{Rel2}$. Note that, despite the
high exposure time, the effective net counts number is of order of 20 per cent
or less in the bins more external than $R_{2500}$ due to the fact of the
background.

In the left panels of Fig.~\ref{fig:photonimg1}, we report the photon images
of our simulated sample. All the images are binned to $2''$ pixels and divided
by their instrument maps. They are computed in the soft [0.5-2 keV] energy
band, to better resolve the presence of small high-emissivity structures or
merging objects that otherwise could be hidden by the background.  The central
panels of Fig.~\ref{fig:photonimg1} present the spectroscopic-like temperature
maps, with the isoflux contours over-plotted.  These maps are directly
produced from the gas particles in the simulation and show the thermal
structure of the systems as would been obtained by ideal X-ray observations of
the simulated clusters.  On the right panels of the same figures, we plot the
histograms of the emission measure, i.e. the sum of the square of the gas
density of all the particles inside the box image for $C_{Hot}$, $C_{Pert}$
and $C_{Merg}$ or inside $R_{2500}$ for $C_{Rel1}$ and $C_{Rel2}$, for each
bin of temperature and per unit volume. The vertical lines represent the
cluster temperature, $T_{2500}$.  We notice that the distribution is quite
regular and Gaussian only for the relaxed cluster $C_{Rel2}$, while many
features are evident for the other systems, revealing their perturbed dynamic
state.  Nevertheless, we notice that $T_{2500}$ is always lower than the value
corresponding to the peak in the emission measure distribution.  This is the
consequence of fitting with a single temperature the spectra of cluster
regions that, as highlighted by the temperature map, show significant
gradients or are highly thermally complex \citep[see,
e.g.][]{2004MNRAS.354...10M}.
 
\subsection {Description of the clusters in the sample} 

 $C_{Hot}$ and $C_{Pert}$ have a quite complex cluster structure.  This is
evident in the emission measure distributions, which are very broad. The fact
that they extend to low temperatures indicates the presence of many cold blobs
of gas.  Some of them can be seen in the soft X-ray images as
bright X-ray blobs distributed around the clusters.  The cool nature of these
X-ray blobs is confirmed by the temperature maps. To estimate the
cluster mass of these complex clusters, we follow the observational approach of
excluding from the spatial and spectral analysis all sub-clumps clearly
detectable from the soft X-ray image.  The excluded regions are shown as green
circles in Fig.~\ref{fig:photonimg1}.  We know that this procedure partially
helps to reduce thermal contamination from colder substructures, however many
thermal structures may still remain as they do not produce detectable
perturbations in the soft X-ray image.

The largest and hottest cluster, $C_{Hot}$, is actually a still forming
object that has not reached the equilibrium state and is undergoing merging
events by small substructures.

$C_{Pert}$ presents, in the southern hemisphere, a sub-clump infalling
toward the cluster centre and merging with a cold structure located at 1.5
arcmin from the centre.

$C_{Merg}$ appears very relaxed, with no evidence of any peculiar 
structure both in the photon image and in the temperature map.  The 
elliptical shape of the isoflux contours is the only imprint of a 
recent major merging happened at $z\approx 0.2$.  A further evidence 
of this recent impact with another big clump comes from the histogram 
of the emission measure that peaks at two different temperatures ($T 
\sim 4.1$ keV and $T \sim 3.4$ keV). 
 
The simulated clusters $C_{Rel1}$ and $C_{Rel2}$ have a regular photon 
image with round isophotes. On the other hand, the temperature map of 
$C_{Rel2}$ shows the presence of a cold arc in the northern part at 
$\approx 1$ arcmin from the centre.

\section{AN OBSERVATIONAL-LIKE X-RAY ANALYSIS}  
  
The event files produced by \xmas have been then processed using the CIAO
3.0.1 software (see http://cxc.harvard.edu/ciao/) and the calibration files in
CALDB 2.23. Surface brightness profiles and spectra have been created and
analyzed as described below. In both analyses we exclude the sub-clumps
detected in the photon image of $C_{Hot}$ and $C_{Pert}$. Moreover, in all
clusters we mask the inner 50 kpc to avoid the influence of the central
cooling part. The excluded regions are marked by green circles in the photon
images.

\subsection{Spatial analysis: brightness profiles}  
 
The surface brightness profiles have been extracted from [0.5-5 keV] 
images that have been corrected thanks to the corresponding 
instrumental maps. 
 
To build the profile we consider annuli centred on the X-ray peak 
(in any case, we verified that the off-set from the actual minimum 
of the cluster potential well is never  larger than 2
arcsec) by  requiring at least 5000 total counts per bin.

\begin{figure} 
\psfig{figure=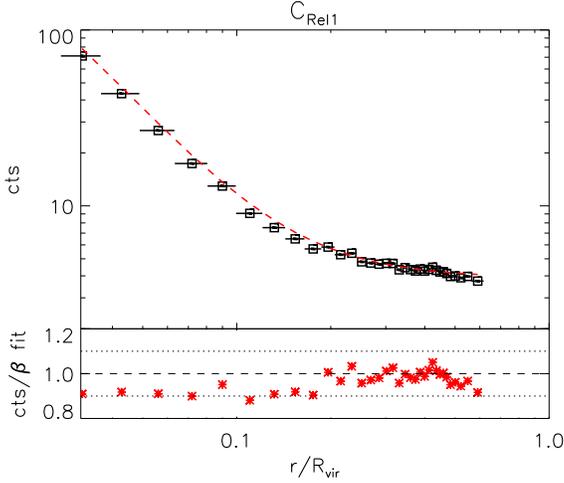 ,width=0.45\textwidth}   
\caption[]{ 
The surface brightness profile (in units of photon counts) of the 
simulated cluster $C_{Rel1}$.  Open squares represent the values 
extracted from the X-ray analysis, the horizontal bars correspond to 
the bin sizes. The dashed curve is the corresponding best-fit 
$\beta$-model, with $\beta=0.52$ and $r_c=21$ kpc (= 0.013 $R_{\rm 
vir}$). The bottom panel shows the ratio between the profile from the 
X-ray analysis and the best-fit $\beta$-model.} 
\label{fig:sb}  
\end{figure}

The azimuthal profile has been fitted  with a $\beta$-model 
\citep{cavaliere1976}: 
\begin{equation}  
S (r) = S_0 \left[1+\left(\frac{r}{r_{\rm c}}\right)^2 \right]^{0.5 
-3\beta} + c_{\rm bkg},
\label{eq:surfb} 
\end{equation}
  where $r_c$ is the core radius, $\beta$ is the power coefficient, and the
constant $c_{\rm bkg}$ is the value added to take into account the background
present in the image.  The best-fit parameters obtained by analyzing our
clusters are listed in Table~\ref{tab:param}. For all our systems, the
deviations between the surface brightness profile and the $\beta$--model fit
are less than 10 per cent in the interval between $0.1 R_{\rm vir}$ and $1
R_{\rm vir}$ (see, e.g., Fig.~\ref{fig:sb} for cluster
$C_{Rel1}$). Nevertheless, the model shows a significant discrepancy in unit
of $\sigma$, as shown by the large values of the reduced $\chi^2$ reported in
Table~\ref{tab:param}. Overall, we conclude that the $\beta-$model cannot
reproduce the profile of the X-ray emission of the simulated clusters and we
find that also a double $\beta$--model is statistically excluded by our data.

\begin{table}  
\caption{ Best-fit parameters used for the X-ray mass estimators and obtained
from the spatial and spectral analysis of the five objects in our sample.
$r_{\rm c}$ and $\beta$ are the core radius (in kpc) and the power coefficient
of the $\beta$--model; $c_{\rm bkg}$ is the constant representing the
background (Eq.~\ref{eq:surfb}); $\gamma$ is the polytropic index;
$\chi^2_{red,\beta}$ and $\chi^2_{red,\gamma}$ are the reduced $\chi^2$
associated to the $\beta$--model and polytropic relation fitting,
respectively; d.o.f. are the degrees of freedom of the fits; $T_{2500}$ is the
spectroscopic temperature calculated inside $R_{2500}$; $\sigma_{2500}$ is the
associated error at 68 per cent confidence level; $\chi^2_{red}$ is the
reduced $\chi^2$, evaluating the goodness of the spectral fit; $r_{\rm s,NFW}$
and $c_{\rm NFW}$ and $r_{\rm s,RTM}$ and $c_{\rm RTM}$ are the scale radius
(in kpc) and the concentration of the NFW and RTM mass model,
respectively.}
 
\begin{tabular}{c c@{\hspace{.7em}} 
c@{\hspace{.7em}} c@{\hspace{.7em}} c@{\hspace{.7em}} 
c@{\hspace{.7em}}} 
\hline \\  
             &$C_{Hot}$      & $C_{Pert}$  & $C_{Merg}$  & $C_{Rel1}$   & $C_{Rel2}$ \\ 
 & & \\  
$r_{\rm c}$  &85$\pm$2       &101$\pm$6    & 61$\pm$2     &21$\pm$1     &14$\pm$1.5 \\ 
$\beta$      &0.56$\pm$0.01  &0.44$\pm$0.01    &0.62$\pm$0.01 &0.52$\pm$0.01&0.56$\pm$0.01 \\ 
$c_{\rm bkg}$& 1.78          &4.30     & 3.91      &3.88         & 4.06\\ 
$\chi^2_{red,\beta}$ (d.o.f.)&4.2 (13)&3.9 (70)& 7.6 (44)&5.3 (39)& 11.7 (29)\\ 
$\gamma$     &1.09$\pm$0.03 &1.11$\pm$0.15&1.10$\pm$0.02 &1.11$\pm$0.04&1.06$\pm$0.03 \\ 
$\chi^2_{red,\gamma}$ (d.o.f.)&1.4 (6)&0.6 (4) & 0.5 (10)&0.9 (5) & 2.6 (3)\\ 
 & & \\  

$T_{2500}$      & 11.18      & 3.87   & 3.76        & 3.39        & 2.64    \\ 
$\sigma_{2500}$ &$\pm0.48$ &$\pm 0.36$  &$\pm0.08$  &$\pm0.12$   &$\pm0.08$\\
$\chi^2_{red}$ (d.o.f.) & 0.51  (45)     &  0.36 (39)   & 0.44  (103)     &0.31 (30)     & 0.37 (80)\\ 
&&\\ 

$r_{\rm s, NFW}$   &668$\pm$47&350$\pm$100&216$\pm$103 & 82$\pm$45 &252$\pm$38\\
$c_{\rm NFW}$      &6.0$\pm$0.3 & 6.1$\pm$2.5 & 8.6$\pm$2.3 & 16.2$\pm$7.9 &7.2$\pm$1.0 \\
$r_{\rm s, RTM}$   &693$\pm$112 & 351$\pm$121 & 148$\pm$69   & 41$\pm$20    & 226$\pm$70  \\
$c_{\rm RTM}$      &7.5$\pm$1.1 & 7.4$\pm$5.3 & 14.3$\pm$5.4   & 36.2$\pm$8 & 9.9$\pm$3.3\\ 
\\ 
\label{tab:param} 
\end{tabular}  
\end{table}  

\subsection{Spectral analysis}  
\label{sec:spectr} 
To measure the overall temperature we extract spectra from circular regions
centred on the cluster centre and with a radius $r=R_{2500}$. To calculate the
temperature profiles we extract spectra from annular regions for which the
net counts (i.e. belonging to the cluster) are greater than 5000 and are
at least 25 per cent or the total counts.  As in the spatial analysis the
regions marked by green circles in Fig.~\ref{fig:photonimg1} are excluded
from the spectral analysis.
 
For each region, the ancillary response file (ARF) and the redistribution
matrix file (RMF), weighted by the X-ray brightness in the [0.3--2 keV] energy
range, are computed by using the CIAO tools {\tt mkwarf} and {\tt mkrmf}.
Source spectra are extracted from the event file, re-binned and analyzed in
the [0.5--6 keV] band. Background spectra are extracted from the background
event file for the same source regions.  A thermal model ({\tt mekal})
absorbed by the Galactic column density is fitted to the data by using the
$\chi^2$ statistic in the {\it XSPEC} package \citep{1996ASPC..101...17A}.
The only free parameters are the gas temperature and the normalization, being
Galactic absorption $N_{\rm H}$, redshift $z$ and metallicity $Z$ fixed to the
input values adopted in the \xmas run: $N_{\rm H} = 5 \times 10^{20}$
cm$^{-2}$, $z=0.175$ and $Z = 0.3 Z_{\odot}$.

In Table~\ref{tab:param}, we quote the best-fit values of the 
spectroscopic temperature, $T_{2500}$, with the corresponding 
68\% confidence level error $\sigma_{2500}$. 
These value are also shown as vertical lines in 
the right panels of Fig.~\ref{fig:photonimg1}.

\subsection{Deprojection results}

To compute the mass through the equation of the hydrostatic equilibrium as
described in the next section, we need to recover the three-dimensional
profiles of the gas temperature and density by deprojecting the quantities
measured in the X-ray spectral analysis. We adopt the deprojection technique
presented in \cite{ettori2002}, that is described briefly here.  For each
annulus with luminosity $L_{\rm ring}$, a single thermal model is fitted to
the projected spectrum as described in the previous section, giving the
temperature $T_{\rm ring}$.  By using the geometrical corrections discussed in
\cite{1983ApJ...272..439K}, the photon-weighted, projected-on-the-sky measured
quantities are converted to the values expected in spherical shells by the
relations:
\begin{equation}   
\begin{array}{l}   
n_{\rm e} = \left[ ({\bf Vol}^T)^{-1} \# (EI/0.82) \right]^{1/2}\ , \\   
\epsilon = ({\bf Vol}^T)^{-1} \# L_{\rm ring} \ , \\   
T = ({\bf Vol}^T)^{-1} \# (L_{\rm ring} T_{\rm ring}) \, / \, \epsilon\ ,  
\end{array}   
\end{equation} 
where we define the electron density $n_{\rm e}$ in terms of proton 
density $n_{\rm p}$ as $n_{\rm e} = n_{\rm p}/0.82$, the Emission 
Integral is $EI = \int n_{\rm e} n_{\rm p} dV = 0.82 \int n_{\rm e}^2 
dV = K \times 4 \pi d_{\rm ang}^2 (1+z)^2 \times 10^{14}$, $K$ is the 
normalization of the {\tt mekal} model, and the symbol $\#$ represents 
the ``matrix product'' operator.  The matrix ${\bf Vol}$ contains the 
values of the fractions of the spherical volume of a shell seen at 
each ring.  [The notation $({\bf Vol}^T)^{-1}$ indicates that the 
matrix is firstly transposed and then inverted].  The outputs are then 
the measurements of the electron density and plasma temperature in 
each volume shell with given inner and outer radii. 
 
\begin{figure*} 
\psfig{figure=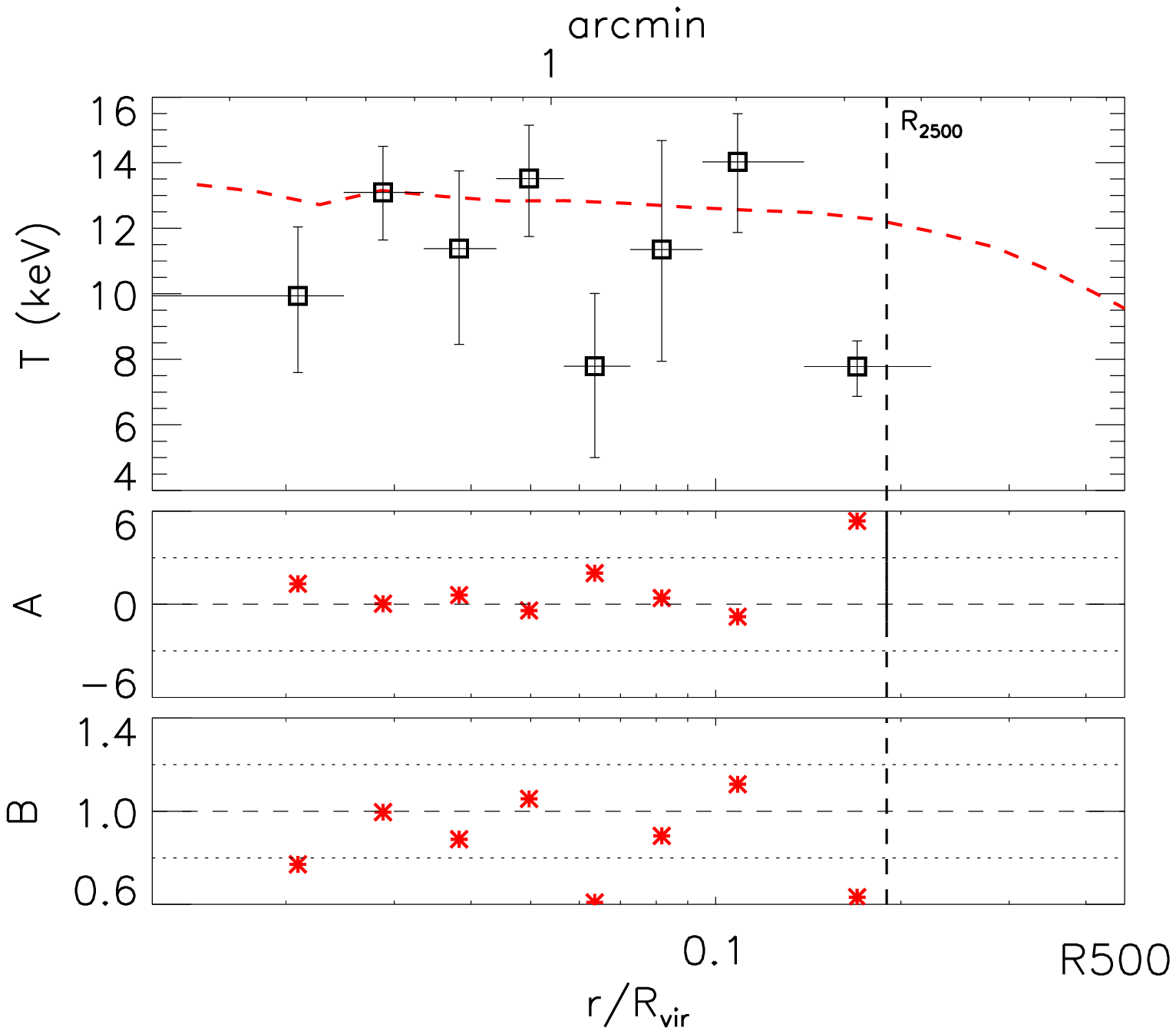,width=0.45\textwidth}  
\psfig{figure=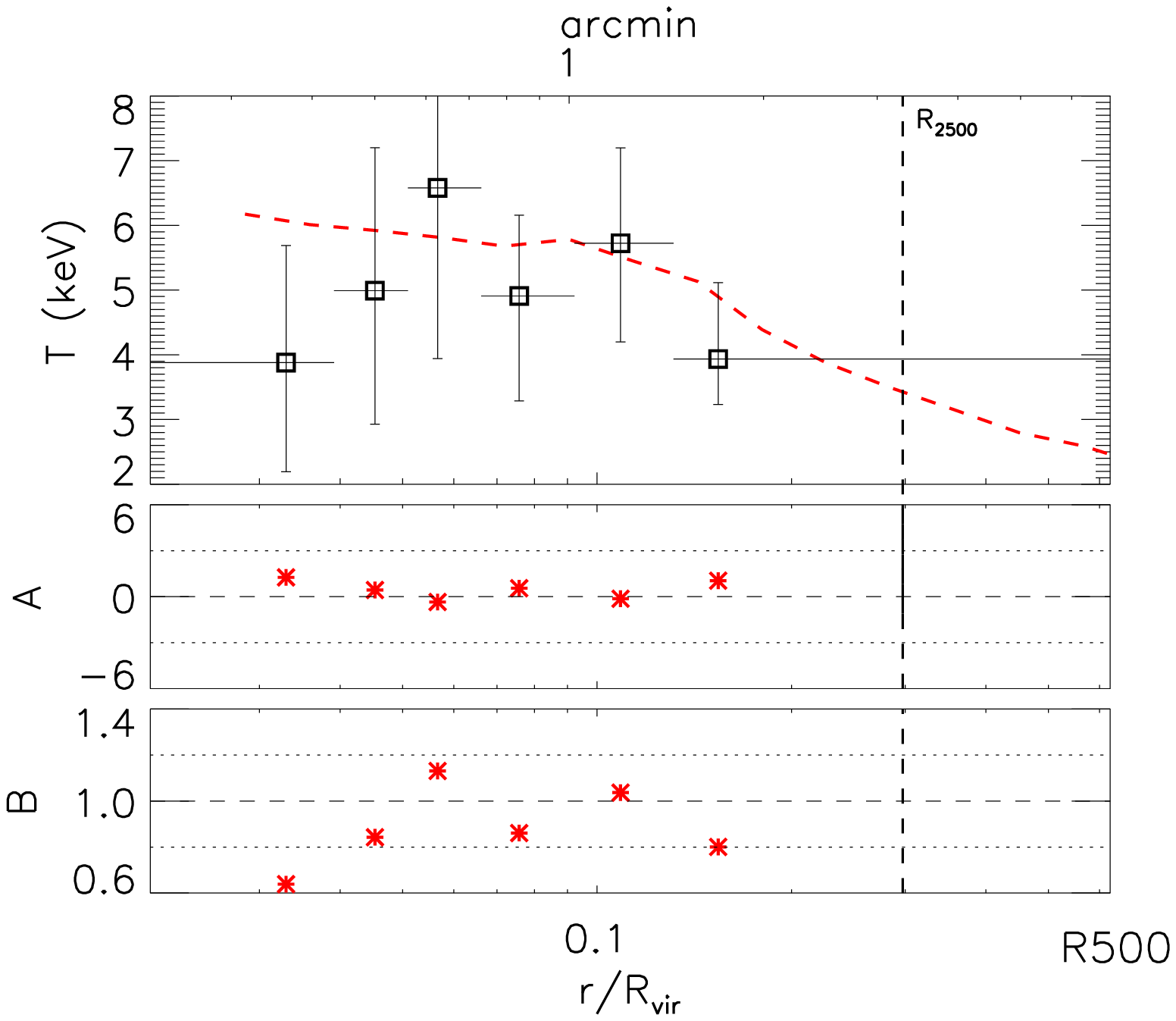,width=0.45\textwidth}  
\psfig{figure=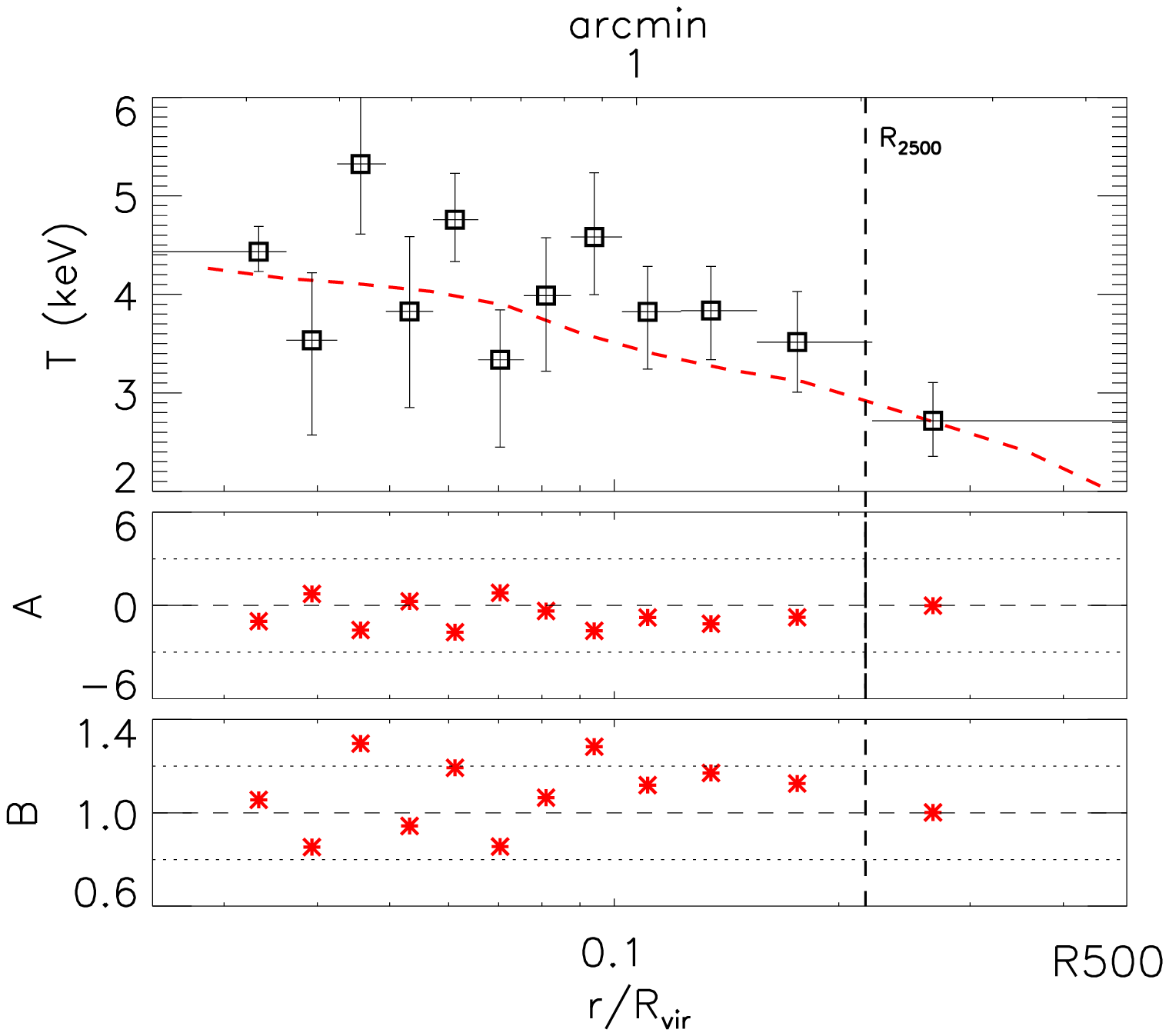,width=0.45\textwidth}  
\psfig{figure=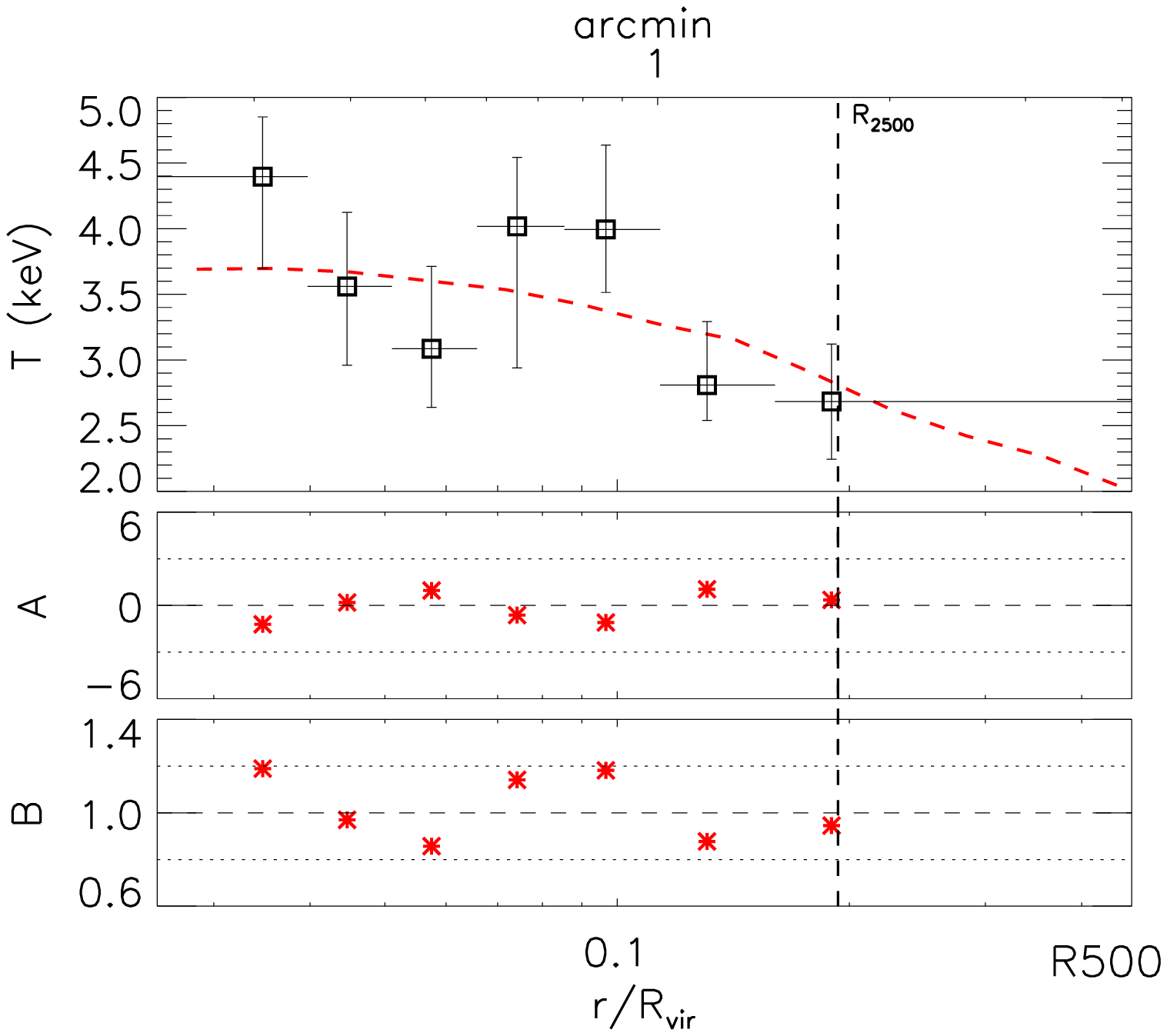,width=0.45\textwidth}  
\psfig{figure=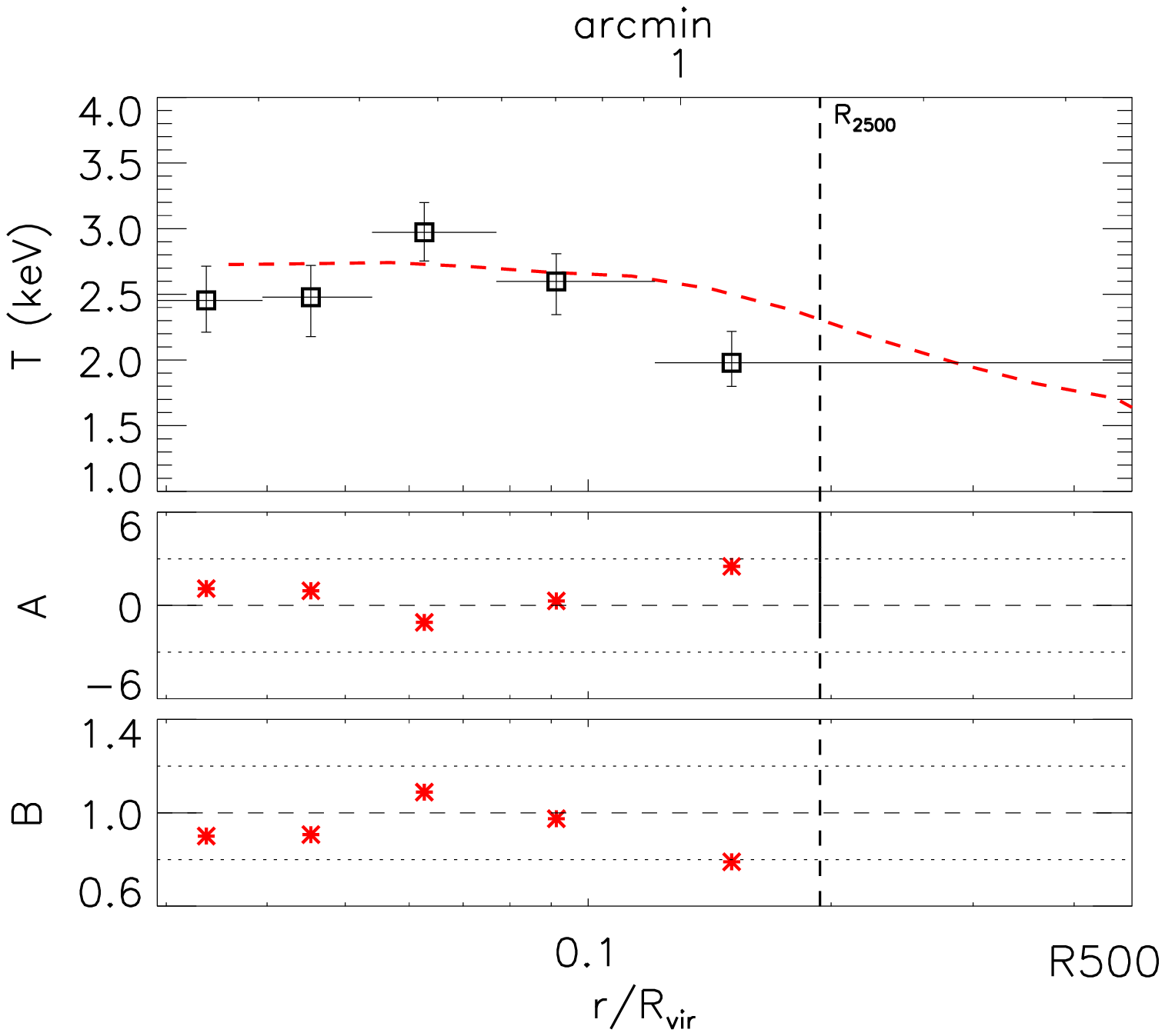,width=0.45\textwidth}  
\caption[] 
{Deprojected temperature profile for $C_{Hot}$, $C_{Pert}$, $C_{Merg}$, $C_{Rel1}$, 
and $C_{Rel2}$.  The dashed line represents the profile for the 
three-dimensional mass-weighted temperature $T_{\rm mw}$ as obtained from the 
hydrodynamic simulation.  The open squares are the values extracted from the 
X-ray analysis, the vertical bars are $1\sigma$ errors ($\sigma_{\rm spec}$), 
while the horizontal ones correspond to the bin sizes. The bottom graphs in 
each panel show quantities related to the differences between the two 
temperatures: $A\equiv(T-T_{\rm mw})/\sigma_{\rm spec}$ and $B\equiv (T/T_{\rm 
mw})$.  The dotted lines indicate $A=(-3,3)$ and $B=(0.8,1.2)$. } 
\label{fig:temp_depr} 
\end{figure*}

We compare in Fig.\ref{fig:temp_depr} the deprojected temperature measure,
$T$, with the three-dimensional mass-weighted estimate, $T_{\rm mw}$,
resulting from the simulations.  The mass-weighted temperature is estimated
directly from the hydrodynamic simulations as $T_{\rm mw} = \int m T dV/\int m
dV$, where $m$ is the mass of each gas particle.  $T_{\rm mw}$ is the proper
temperature value we should use in the hydrostatic equilibrium equation to
derive the mass \citep[][ see also Section 4.4]{math_evr01}.  Since the
clusters in our sample have an azimuthally quite symmetric thermal structure,
we find that their $T_{\rm mw}$ profile is not very dissimilar from the
$T_{\rm spec}$ one, even if in the outer regions the values obtained in the
X-ray analysis are systematically lower.  A more quantitative comparison can
be done by looking at the bottom graphs of each panel, where we show a
parameter similar to the one defined in Section~\ref{sec:spectr}:
$A\equiv(T-T_{\rm mw})/\sigma_{\rm spec}$, where now the temperatures are
three-dimensional quantities, and the ratio between the deprojected
temperature and the mass-weighted temperature: $B\equiv (T/T_{\rm mw})$.  We
find that, while $|A| \la 3$ up to $R_{500}$ for all the objects, $B$
indicates that the spectroscopic temperature is within 20 per cent of $T_{\rm
mw}$, with implications on the mass estimates \citep[][ see Section
5.1]{math_evr01, vikh05mass}.
 
\begin{figure} 
\psfig{figure=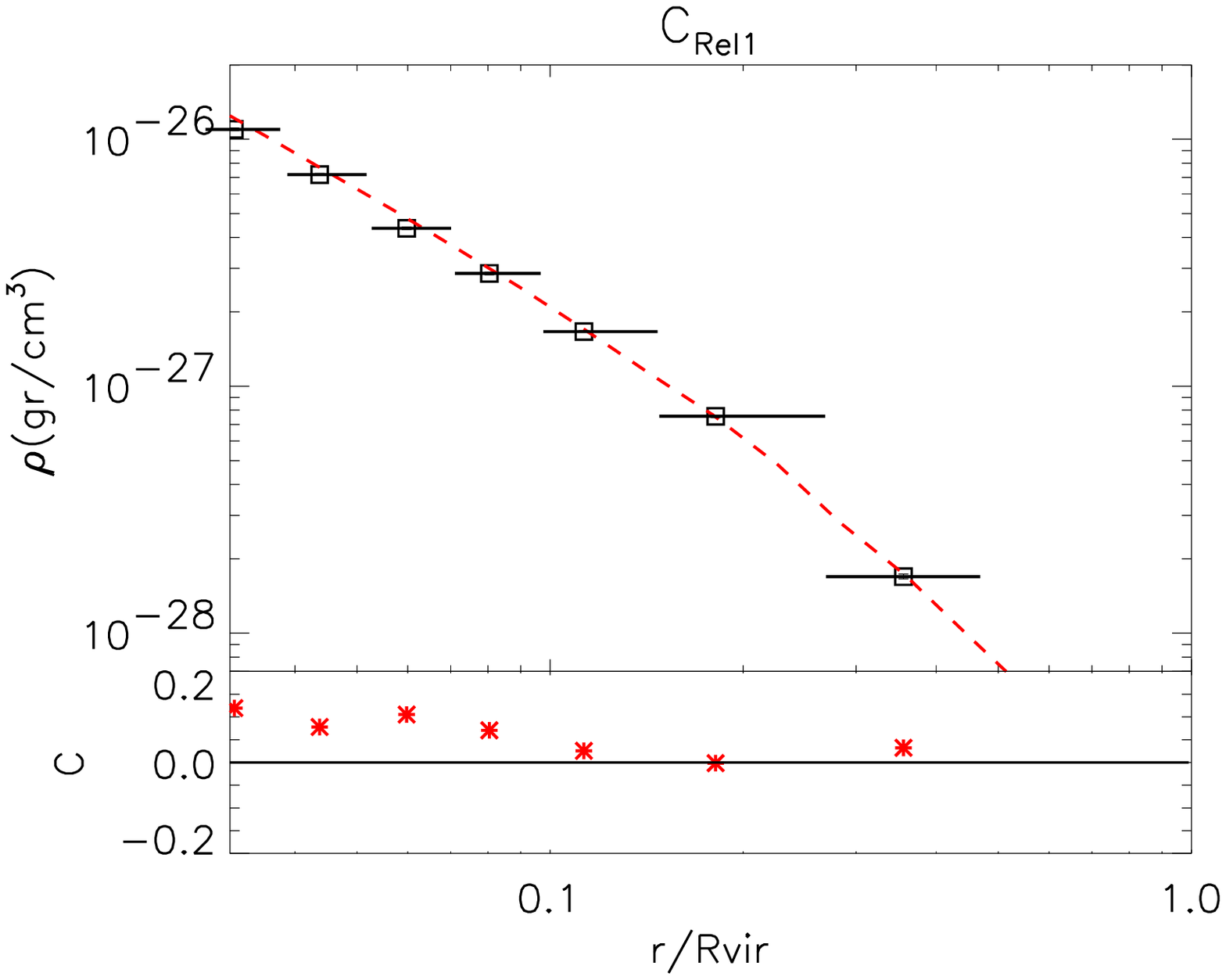,width=0.45\textwidth} 
\caption[] {Deprojected density profile of $C_{Rel1}$. The red dashed line 
represents the density profile as obtained from the hydrodynamic 
simulation ($\rho_{\rm sim}$). The dots are the values extracted from 
the X-ray analysis ($\rho$), the vertical bars (having sizes 
comparable to the dots) are $1\sigma$ errors, while the horizontal 
ones correspond to the bin sizes.  The bottom panel shows the relative 
differences between the true values of the simulation and the derived 
ones, $C \equiv (\rho_{\rm sim} -\rho)/\rho_{\rm sim}$.  } 
\label{fig:den_depr} 
\end{figure}

 Just as an example, in Fig.~\ref{fig:den_depr} we compared to the true
profile $\rho_{\rm sim}$, of the simulated cluster $C_{Rel1}$ (dashed line),
with the gas density obtained from the deprojection technique used.  We find a
good agreement for $r/R_{\rm vir}>0.1$, result that holds also for the other
clusters.
 
 
\subsection{X-ray estimates of the gravitational mass}  
\label{sec:mass} 
 
The ``true'' mass profile, $M_{\rm sim}(<r)$, of the simulated 
objects, obtained by summing all the masses of the particles inside a 
sphere of radius $r$, can be now compared to several different X-ray 
mass estimators, $M_{\rm est}$, that we discuss below. 
 
\begin{itemize}  
\item $M_{\rm est}$ {\it from the  
direct application of the hydrostatic equilibrium equation}: $M_{\rm 
HE}$ 
 
This method \citep[as discussed in][]{ettori2002} makes use of the 
hydrostatic equilibrium (HE) equation under the assumption of 
spherical symmetry to estimate the total gravitating mass: 
\begin{equation}  
M_{\rm HE}(<x) = -\frac{kT(x) x R_{\rm vir}}{G \mu m_{\rm p}}   
\left( \frac{d \ln \rho}{d \ln x} + \frac{d \ln T} {d \ln x} \right) \ ,  
\label{eq:hydr}  
\end{equation}  
where $G$ is the gravitational constant, $\mu$=0.59 is the mean molecular 
weight in a.m.u., $m_{\rm p}$ is the proton mass, $k$ is the Boltzmann 
constant, $x\equiv r/R_{\rm vir}$ and $T$ and $\rho$ are the deprojected 3D 
profiles. The equation is applied directly to the deprojected data by 
estimating the derivatives with respect to $\ln x$ of $\ln(\rho \times T)$ 
without any further smoothing of the measured data points. 
The HE equation is based on simple assumptions: sphericity, static 
gravitational potential and isotropic velocity field.  Recent studies 
have shown how the latter hypothesis is often not satisfied inside 
both simulated clusters 
\citep[see, e.g., ][]{2004MNRAS.351..237R,2004MNRAS.355.1091K}  
and observed ones 
\citep[see, e.g., ][]{zhang2005,dupke2005}.  
\cite{2004MNRAS.351..237R} suggested that in order to correctly compute  
the total mass, the gas velocity has to be taken into account, because 
it is a relevant component of the total energy equipartition between 
gas and dark matter.  The more general model describing the {\it gas 
dynamical equilibrium} can be then written as: 
\ba 
M_{HE,v}(<x)&& =  -\frac{kT(x) x R_{\rm vir}}{G \mu m_{\rm p}} \left[ \frac{d \ln \rho}{d \ln x} 
                +\frac{d \ln T}{d \ln x} \right] \nn \\ 
  & &  -\frac{\sigma_r^2 x R_{\rm vir}}{G} \left[ \frac{d \ln \rho}{d \ln x}  
               +\frac{d \ln \sigma_r^2}{d \ln x} + 2\beta_v(x) \right],  
\label{eq:mxv} 
\ea where $\beta_v (\equiv 1- \sigma_t^2/2\sigma_r^2)$ is the gas velocity
anisotropy parameter and $\sigma_r$ and $\sigma_t$ are the gas radial and
tangential velocity dispersions, respectively.  Since some years, velocity
fields have been studied in simulated cluster as bulk motion \citep[see,
e.g.][]{bn99} or turbulent motions \citep[see, e.g.][]{klaus05,fra06}.
Nevertheless, robust observational measurements of $\beta_v$ and, more in
general, of the gas bulk motions are difficult \citep[see, e.g.,][]{dupke2005}
and due to the failure of {\it Suzaku} satellite spectrometer, we have to wait
the new generation of high spectroscopic resolution X-satellites to determine
the cluster velocity structure. In the present analysis, we evaluate the
second part of Eq.~\ref{eq:mxv} directly from the simulations.
 
\item $M_{\rm est}$ {\it  using $\beta-$model  
and polytropic temperature profile}: $M_{\beta,\gamma}$ 
 
Assuming that the gas density profile is described by a 
$\beta-$model and that it is related to the temperature through the 
polytropic relation ($T \propto \rho ^{\gamma -1}$, with $1 \le \gamma 
\le 5/3$), Eq.~\ref{eq:hydr} can be written as  
\citep[see, e.g., ][]{henriksen1986}: 
\begin{eqnarray}  
M_{\beta,\gamma} (<x) & = & -\frac{kT(x) \ x}{G \mu m_{\rm p}} \left( 
\frac{d \ln \rho}{d \ln x} + \frac{d \ln T}{d   
\ln x}\right)  \nonumber \\ 
 & = & \frac{3 \ \beta \gamma \ T_0 \ r_{\rm c}}{G \mu m_{\rm p}}  
\frac{x_{\rm c}^3}{ (1+x_{\rm c}^2)^{\alpha} } \nonumber \\  
 & = & \frac{0.757 \times 10^{14}}{\mu} \beta \gamma T_0 r_{\rm c}   
\frac{x_{\rm c}^3}{ (1+x_{\rm c}^2)^{\alpha} } \ M_{\odot},  
\label{eq:poly}  
\end{eqnarray}  
where $x_{\rm c}\equiv r/r_{\rm c}$, $r_{\rm c}$ is the core radius (in units 
of $h_{70}^{-1}$ Mpc), $\alpha = 1.5 \beta (\gamma -1) +1$ and $T_0$ is the 
central temperature in keV. 
 
\item $M_{est}$ {\it from an isothermal $\beta-$model}: $M_{\beta}$ 
 
Rather than relating the gas density and temperature accordingly to 
the polytropic law, an isothermal gas is here assumed.  The total mass 
is then obtained from Eq.~\ref{eq:poly} by imposing $\gamma=1$ and 
$T_0=T_{500}$ (reported in Table~\ref{tab:param}). 
 
\item $M_{est}$ {\it through analytic mass models}: $M_{\rm NFW}$ and 
$M_{\rm RTM}$ 
 
This method, widely adopted in the reconstruction of the X-ray mass profile 
\citep[see, e.g., ][]{allen2002},  
makes use of functional forms of the radial mass distribution obtained 
from numerical simulations.  The derived gravitational potential is 
combined with the deprojected gas density profile to recover a 
temperature profile through the numerical inversion of the HE 
equation.  A $\chi^2$ distribution is then obtained by a grid of 
values for the two free parameters of the mass model (the scale radius 
$r_{\rm s}$ and the concentration parameter $c$) by estimating the 
deviation between the numerical temperature profile and the 
deprojected one.  The errors on the best-fit parameters are inferred 
from the distribution of the $\chi^2$ values. 
 
As functional forms of the mass profile, we have used here those 
proposed by \cite{1997ApJ...490..493N} (hereafter NFW) and by 
\cite{2004MNRAS.351..237R} (hereafter RTM), which read 
\ba 
M_{\rm NFW}(<x) & \propto & \log(1 - x \, c_{\rm NFW})- 
\frac{x \, c_{\rm NFW}} {1 - x \, c_{\rm NFW}^3} \ , \nonumber \\  
M_{\rm RTM}(<x) & \propto & 
\left[\frac{(x \, c_{\rm RTM}+2)} {x \, c_{\rm RTM} + 1}^{1/2} -2 \right], 
\label{eq:rtm} 
\ea 
respectively, with $x \equiv r/R_{\rm vir}$.

The values of the corresponding best-fit parameters for our set of 
simulated clusters are reported in Table~\ref{tab:param}. 
\end{itemize}

\section{Results on the mass profiles} 
 
\begin{table}  
\caption{ Deviations between the masses calculated by using the different mass
estimators $M_{\rm est}$ (see their description in the text) and the true mass
$M_{\rm sim}$ as directly obtained from the simulated clusters, in units of
the estimators errors: $|M_{\rm est}-M_{\rm sim}|/\sigma_{\rm est}$.  In the
case of discrepancies greater than 1 $\sigma_{\rm est}$ we report in
parenthesis the value of the percentage difference: $(|M_{\rm est}-M_{\rm
sim}|/M_{\rm est}) \times 100$. The values are computed at $R_{2500}$ and
extrapolated to $R_{500}$ for $M_{\beta}$, $M_{\beta,\gamma}$, $M_{\rm RTM}$,
and $M_{\rm NFW}$.}
\begin{tabular}{  
 c@{\hspace{.8em}} c@{\hspace{.8em}} c@{\hspace{.8em}} c@{\hspace{.8em}} 
 c@{\hspace{.8em}} c@{\hspace{.8em}} } 
\hline \\  
&$C_{Hot}$  & $C_{Pert}$& $C_{Merg}$& $C_{Rel1}$ &  $C_{Rel2}$\\ 
\\ 
 & \multicolumn{5}{c}{${\bf R_{2500}}$}  \\ 
$M_{HE}$            &  1          & 1           & 1             & 1        &  1  \\ 
$M_{HE,v}$          &  1          & 1           & 1             & 1        &  1 \\ 
$M_{\beta,\gamma}$  & $>10$ (47)& $>10$ (49) & $>10$ (23) & $>10$ (37) & $>10$ (36)  \\ 
$M_{\beta}$         & $>10$ (43)& $>10$ (58) & $>10$ (24) & $>10$ (35) & $>10$ (31) \\ 
$M_{\rm NFW}$       &  1          & 1           & 1             &  2 (20) &  3 (24)  \\ 
$M_{\rm RTM}$       &  2 (13)   & 1           & 1             &  2 (20) &  5 (32) \\ 
  & & & & & \\ 
\\ 
 & \multicolumn{5}{c}{${\bf R_{500}}$}  \\ 
$M_{\beta,\gamma}$  &$>10$ (52)&4 (45) &7 (28)  &9 (42)   &7(37) \\  
$M_{\beta}$         & 10   (44)&10 (51)&10 (17) &>10 (31) &>10(25) \\  
$M_{\rm NFW}$       & 1        &1      &1       &2 (35)   &3(46) \\  
$M_{\rm RTM}$       & 3    (56)&1      &1       &3 (27)   &4(81) \\  
\label{tab:mass} 
\end{tabular}  
\end{table}  
 
\begin{figure*}  
\psfig{figure=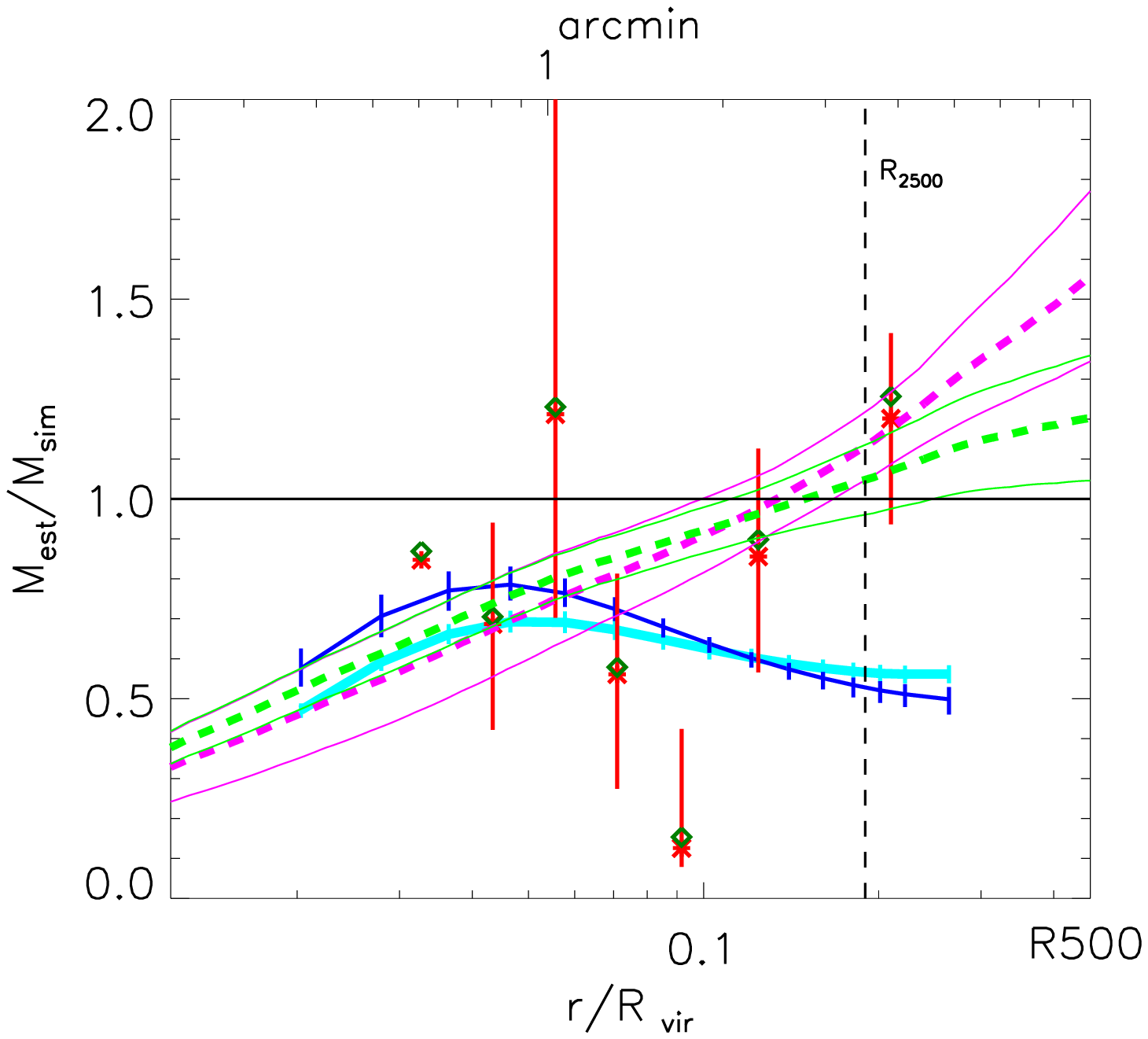 ,width=0.45\textwidth} 
\psfig{figure=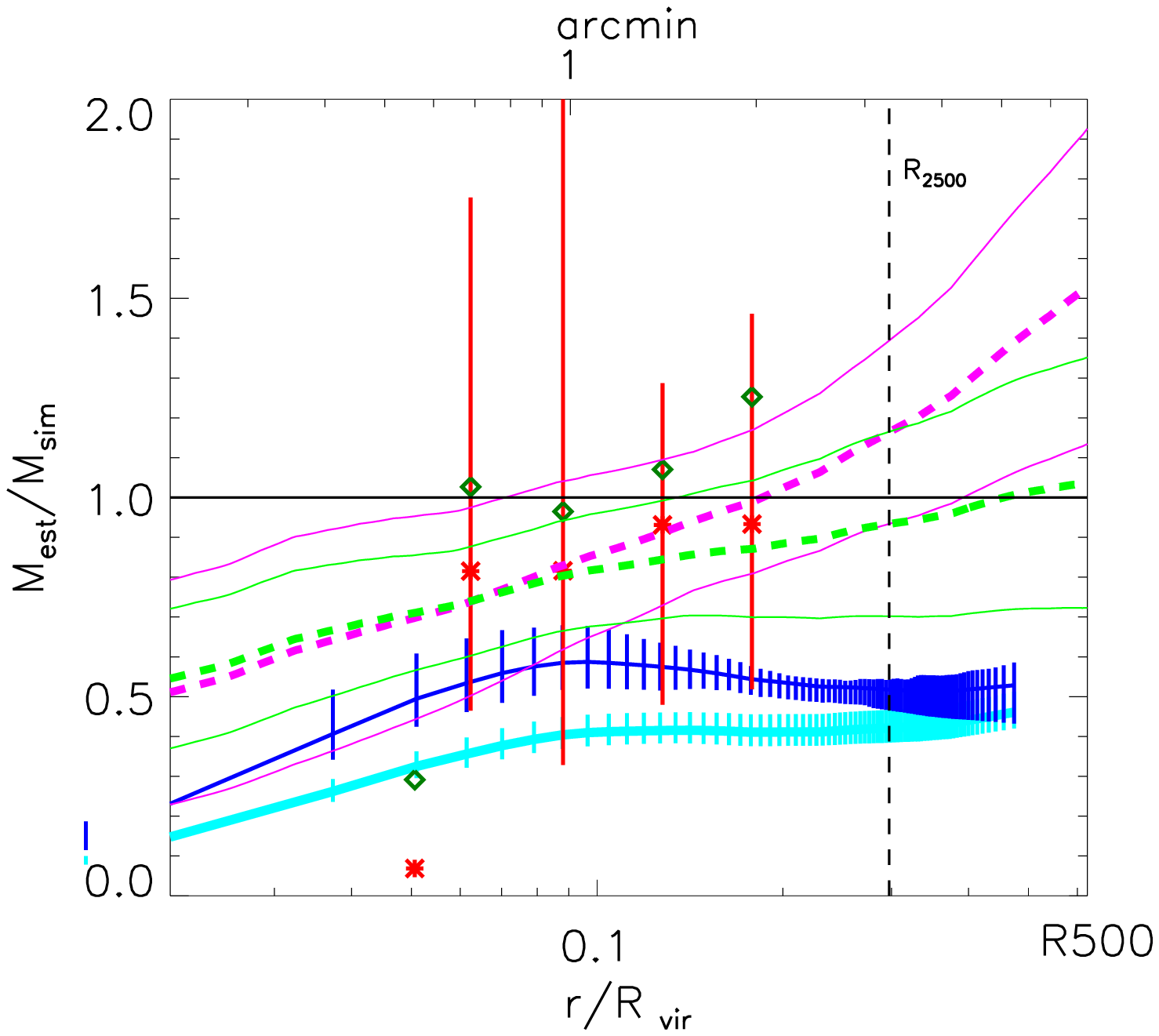,width=0.45\textwidth}  
\psfig{figure=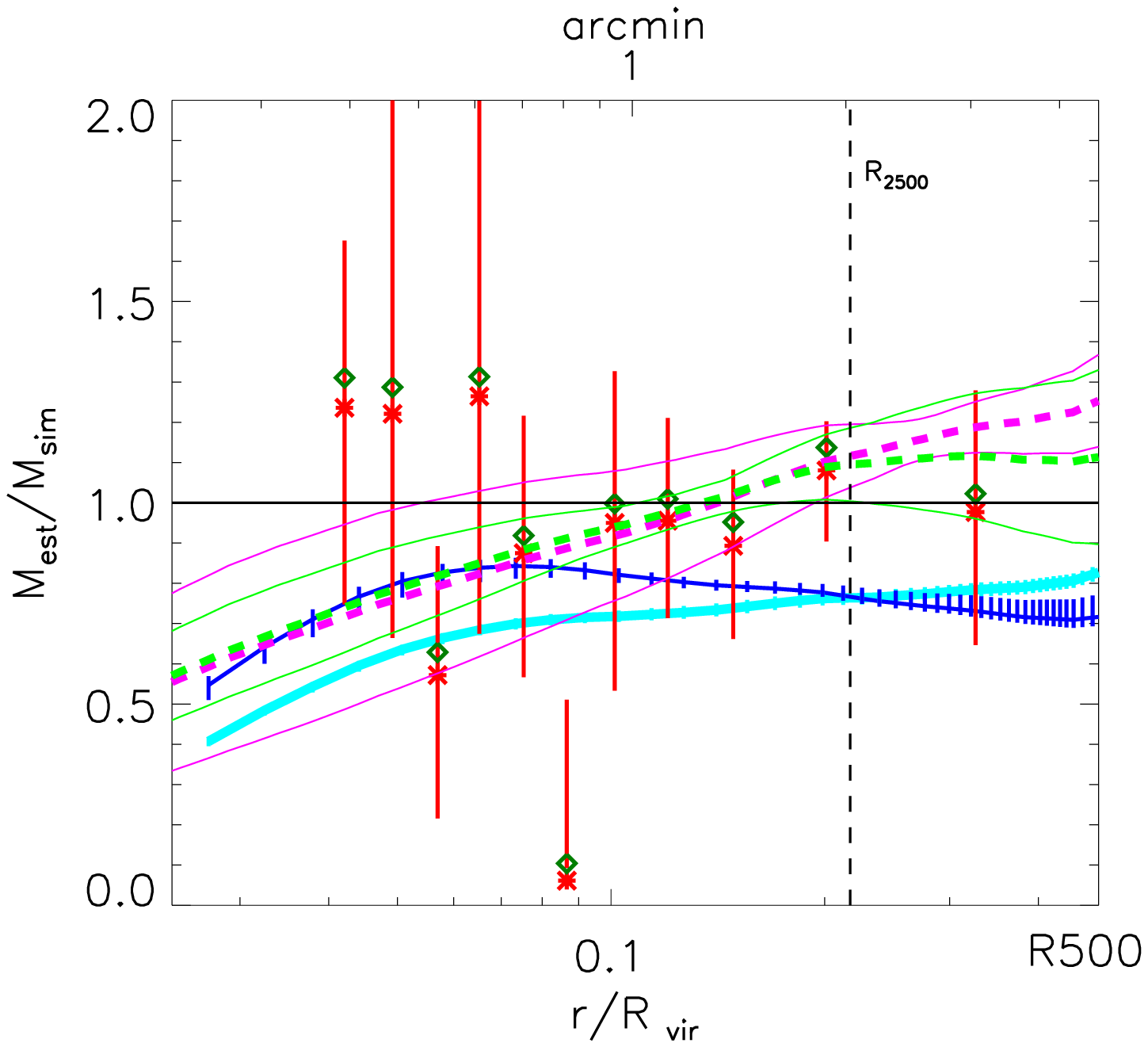,width=0.45\textwidth}   
\psfig{figure=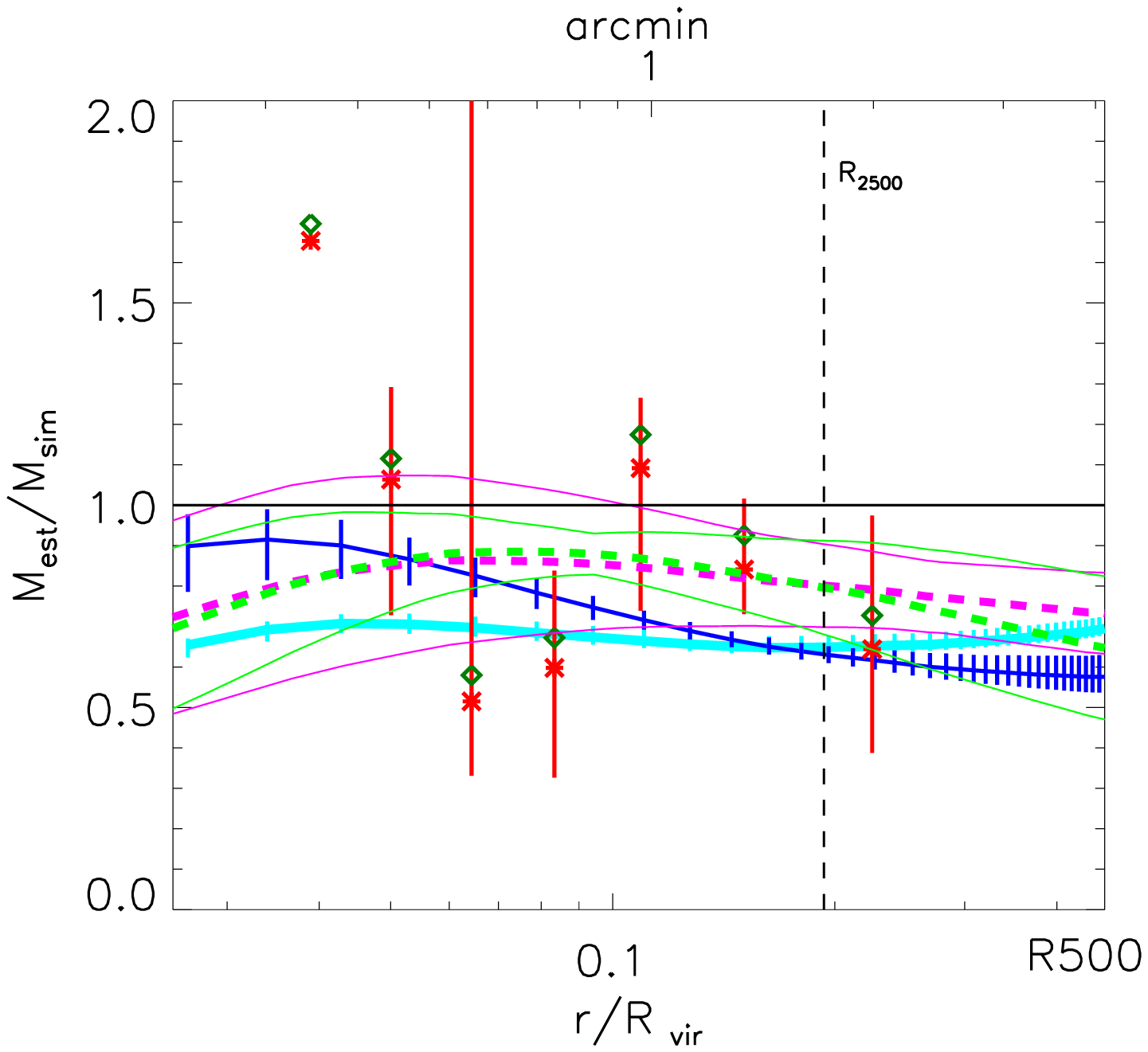 ,width=0.45\textwidth} 
\psfig{figure=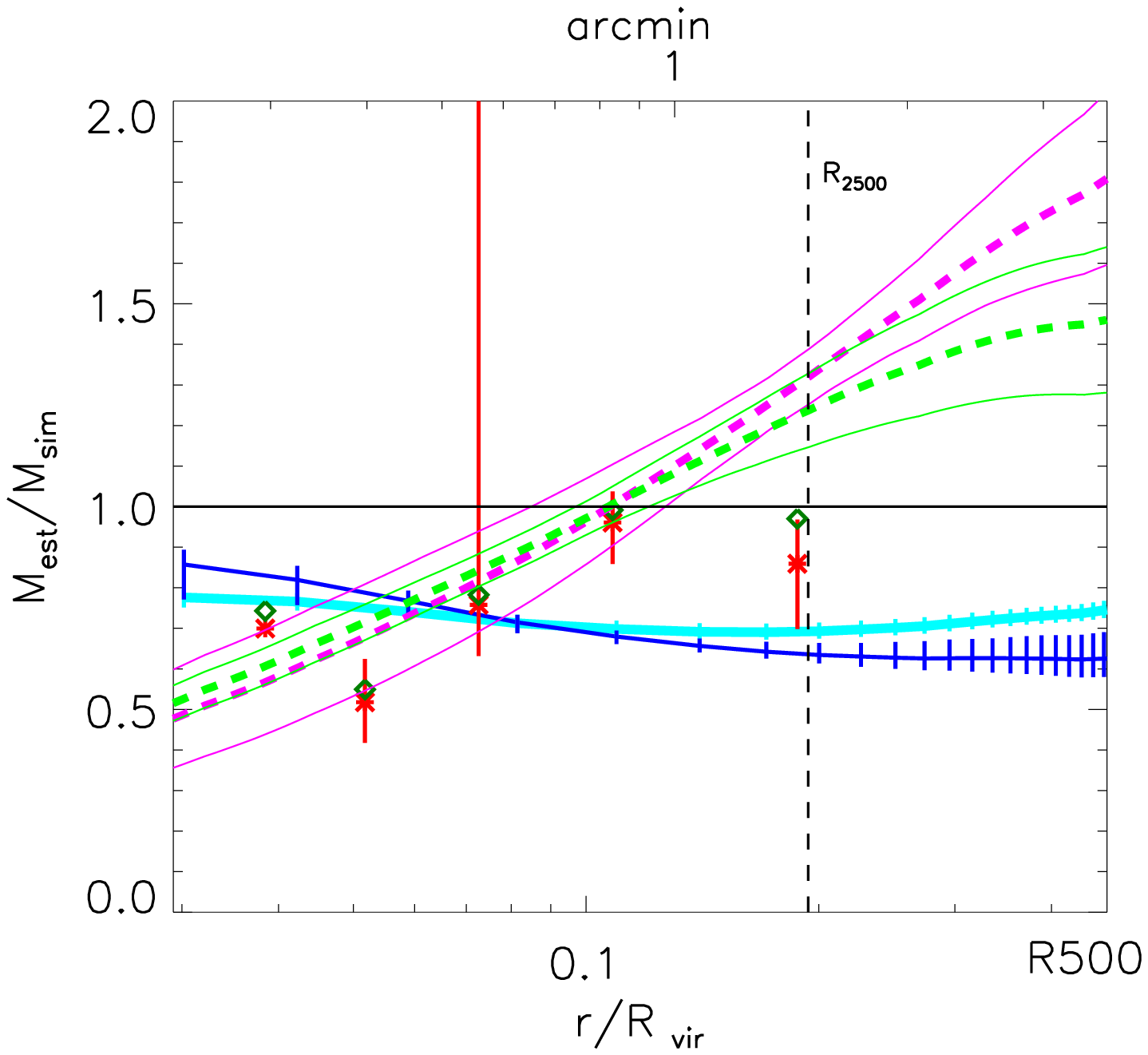 ,width=0.45\textwidth} 
\psfig{figure=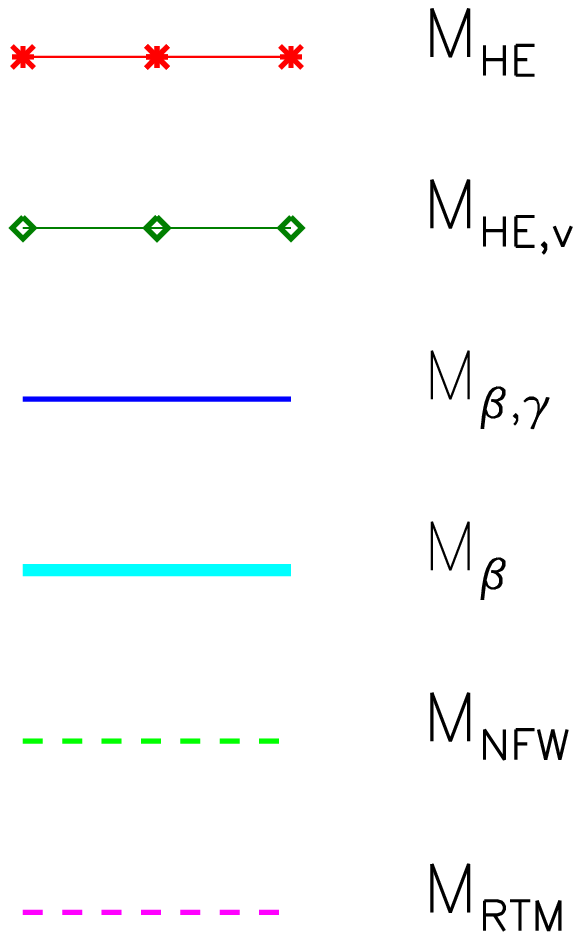 ,width=0.45\textwidth} 
\caption[] {Ratios between the mass profiles derived from the X-ray analysis,
$M_{\rm est}$, and the true mass profile of the simulated cluster, $M_{\rm
sim}$ for our galaxy clusters: $C_{Hot}$, $C_{Pert}$, $C_{Merg}$, $C_{Rel1}$
and $C_{Rel2}$.  The vertical lines indicate $R_{2500}$.  The red asterisks
and the green diamonds represent the mass derived by assuming the hydrostatic
equilibrium, $M_{HE}$ (Eq.~\ref{eq:hydr}) and $M_{HE,v}$ (Eq.~\ref{eq:mxv}),
respectively.  The solid blue and the thick cyan lines refer to the
$\beta-$model, $M_{\beta}$, and $\beta-$model plus polytropic relation,
$M_{\beta,\gamma}$ (Eq.\ref{eq:poly}), respectively.  The green and magenta
dashed lines show the masses derived by assuming the analytic profiles,
$M_{\rm NFW}$ and $M_{\rm RTM}$, respectively (Eq.\ref{eq:rtm}).  The measured
errors (1 $\sigma$ ) on the mass estimators are represented by vertical error
bars or thin solid lines in the case of $M_{\rm NFW}$ and $M_{\rm RTM}$.  }
\label{fig:mass} \end{figure*}   
 
The ratios between the mass profiles reconstructed by using all these methods,
$M_{\rm est}$, and the true mass profiles as extracted by simulated clusters,
$M_{\rm sim}$, are shown in Fig.~\ref{fig:mass}.  Note that the uncertainties
relative to the X-ray mass measurements as shown in Fig.~\ref{fig:mass}
decrease indeed in relation to the dynamic state of the cluster, with the
largest errors being associated to the first three objects ($C_{Hot}$,
$C_{Pert}$, $C_{Merg}$) and the smallest ones to the most relaxed systems,
$C_{Rel1}$ and $C_{Rel2}$.  The vertical dashed lines refer to the radii
corresponding to $R_{2500}$.  In Table~\ref{tab:mass} we quote the differences
between the true mass and the mass estimates in units of the error at this
radius and at $R_{500}$, where $M_{\beta}$, $M_{\beta,\gamma}$, $M_{\rm RTM}$,
and $M_{\rm NFW}$ need to be extrapolated. In the case of discrepancies
greater than 1 $\sigma$ error we report also the percentage difference in
parenthesis.
 
We summarize here our results and, where needed, discuss the main reasons for
the observed discrepancies between the estimated X-ray mass and the total mass
distribution.
 
\begin{itemize} 

\item The mass profile obtained from the HE equation, $M_{HE}$, dependent upon
the measured temperature profile (see Fig.~\ref{fig:temp_depr}).  Large errors
on, and irregular radial distribution of, the temperature values induce large
scatter on the reconstructed $M_{HE}$ measurements.  As clear from the figure,
this mass estimator produces large errors mainly related to the background
level. For this reason most of these measurements are within $1 \sigma$ from
the expected values, even in objects that are still undergone to some (minor)
mergers and for which the hydrostatic equilibrium might have not been reached
yet ($C_{Hot}$ and $C_{Pert}$).  If we focus only on the best fit value, we
notice that, in the case of $C_{Hot}$, $M_{HE}$ overestimates the true mass
due to the steepness of the temperature profile at $R_{2500}$ owing to the
wrong measurement of temperature profile in the last bin. On the other hand,
it underestimates the true mass in relaxed objects (see Fig.~\ref{fig:mass}).

The effect of the gas bulk motions are always smaller 10 per cent, excluding
for the cluster $C_{Pert}$. In this case the kinetic energy is significant
mostly because of the entering merging blob in the South and can still
contribute between 20 and 30 per cent to the total mass measurements. The
neglected kinetic pressure term can partially compensate for the present 
best fit deviations in relaxed systems.

\item The analytic masses, $M_{\rm NFW}$ and $M_{\rm RTM}$, are reconstructed
starting from the hydrostatic equilibrium equation, therefore they follow the
behaviour of the $M_{HE}$ profile, but they have smaller errors.
Nevertheless, we notice that the statistical errors are still too high to show
significant mass discrepancies (Fig.~\ref{fig:mass}).

\item  The $\beta$--model  gives the worse mass reconstruction
with mass discrepancies $>10\sigma$ significant \citep[see
also][]{bartelmann1996,muanwong2002,2004MNRAS.351..237R,borgani2004,
rasia.etal.05}.  $M_{\beta,\gamma}$ is 5-10 per cent lower than $M_{\beta}$ in
the outer regions ($r> R_{500}$) owing to the use of a polytropic law to
describe the relation between the measured temperature and density profiles.
Here we note that, in our simulated clusters, the correlation between $\ln
T(r)$ and $\ln \rho(r)$ cannot be reproduced by a single linear relation and
requires the polytropic index $\gamma= 1+ \ln T /\ln \rho$ to vary between 1.1
and 1.2 moving outwards.  The use of an inappropriate assumption on the
functional form of the temperature profile propagates to the mass measurements
making them less accurate than the estimates done with the isothermal
$\beta$--model.  The main sources of the difference for these models are (i)
the poorness of the $\beta$--model in describing the density profile, (ii) the
incorrect assumption of isothermality or above all of the polytropic relation,
and (iii) the uncertain determination of the $\beta$--model parameters (which
we will discuss in the next subsection).

\end{itemize} 
 
Looking at the behaviour of each single cluster, we notice that,
even for the dynamically disturbed objects, the central values
of the total mass can be recovered within 20-30 per cent at 
$R_{2500}$, the situation being worse at $R_{500}$.

\subsection{On the NFW concentration and $\beta$ parameter} 
 
The computation of the mass directly depends on the values of the best-fit 
parameters, like the NFW concentration, $c_{\rm NFW}$, or the $\beta$ 
value in the $\beta-$model.  The former is evaluated from the mass 
functional form that better reproduces the observed temperature 
profile at $r >50$ kpc.  The inner 50 kpc are also excluded in 
the spatial analysis that provides an estimate of $\beta$, to avoid 
any influence from the central cooling region where a too large amount 
of cool gas concentrates in our simulated clusters 
\citep[see, e.g., ][]{borgani2004}. 
 
We investigate here how the measurement of these parameters depends upon the
considered radial range.  When we also include in the analysis the central
region within 50 kpc, the concentration parameter becomes larger, but the
reduced $\chi^2$ also increases.  The cluster $C_{Merg}$ is an example:
$c_{\rm NFW}$ rises from 8.6 to 28 and the reduced $\chi^2$ from 0.9 to 2.6.
On the contrary, the value for $\beta$ does not vary significantly (the most
relevant deviation is for $C_{Pert}$, where $\beta$ changes from 0.56 to
0.47).  Also in the spatial fit of the surface brightness, when including the
inner 50 kpc region the reduced $\chi^2$ of the fit increases.  While this
systematic effect can be easily handled from an observational point of view,
the influence of the outer end of the radial range is limited by the field of
view of the detector and by the background.  From a direct fit of the data
from the simulations, we find that both parameters depend strongly upon the
adopted outer radius, and that, while $c_{\rm NFW}$ decreases, $\beta$
increases when larger radii are considered \citep[see
also][]{NFW1995.1,bartelmann1996,borgani2004}.  The same trend occurs when we
fit an NFW and a $\beta-$model to the density profile. For example, we measure
$c_{\rm NFW}\approx 5.3$ by fitting the simulated density profile of
$C_{Merg}$ out to the virial radius (see Table~\ref{tab:cl}), while $c_{\rm
NFW}\approx 7.0$ if we limit the fit to the region mapped in the \chandra
field-of-view and $c_{\rm NFW}\approx 8.6$ if we use the deprojected
temperature profile.
 
In summary, the concentration parameter and the $\beta$ value are 
affected by both the method used to derive them and the radial interval 
over which they are measured.  These two limitations have the same 
effect: the observed $c_{\rm NFW}$ and $\beta$ will be always higher 
and lower, respectively, than the values directly derived by fitting 
the density of simulated clusters up to the virial radius.

\subsection{Effect of the background level}

\begin{table}
\caption{{ The same of Table~\ref{tab:mass} but derived using a background level   100 lower that the \chandra value (see text in Sect.5.2).}}

\begin{tabular}{  
 c@{\hspace{.8em}} c@{\hspace{.8em}} c@{\hspace{.8em}} c@{\hspace{.8em}} 
 c@{\hspace{.8em}} c@{\hspace{.8em}} } 
\hline \\  
&$C_{Hot}$  & $C_{Pert}$& $C_{Merg}$& $C_{Rel1}$ &  $C_{Rel2}$\\ 
\\ 
 & \multicolumn{5}{c}{${\bf R_{2500}}$}  \\ 
$M_{HE}$           & 1        &  1           & 7 (30)&$>10$ (30)&$>10$ (28)  \\ 
$M_{HE,v}$         & 1        &  1           & 6 (24)& 8 (22)   &9 (17)\\ 
$M_{\beta,\gamma}$ &$>10$ (46)&$>10$(33)& $>10$ (26)&$>10$ (40)&$>10$ (37)  \\ 
$M_{\beta}$        &$>10$ (44)&8 (30)    &$>10$ (16)&$>10$ (31)&$>10$ (26)\\ 
$M_{\rm NFW}$      &8     (19)&4 (34)    &$>10$ (27)&$>10$ (13)& 5(10)  \\ 
$M_{\rm RTM}$      &5     (18)&5 (31)    &$>10$ (19)& 6 (16)& 4 (11)  \\ 
  & & & & & \\ 
\\ 
 & \multicolumn{5}{c}{${\bf R_{500}}$}  \\ 
$M_{HE}$           &1          &1         &5           &4 (15)    &9 (32)   \\  
$M_{HE,v}$         &2 (30)   &1         &3           &1           &4 (13)  \\  
$M_{\beta,\gamma}$ &$>10$(44)&11 (35) &$>10$ (30)&$>10$ (45)&$>10$ (37)\\  
$M_{\beta}$        &$>10$(42)& 2 (9)  &6  (7)    &$>10$ (25)&$>10$ (19)\\  
$M_{\rm NFW}$      &2 (10)   & 3 (40) &$>10$ (46)&7  (7)    &3 (13)  \\  
$M_{\rm RTM}$      &1         & 5 (30) &$>10$ (24)&3  (6)    &4 (8)  \\  

\label{tab:mass2}
\end{tabular}
\end{table}

In the previous analysis, we follow an observational approach to recover the
mass from \chandra observations.  Mainly due to the relatively high background
level of \chandra, this results in statical errors for the temperature profile
that, in some cases, are too large to be able to show significant mass
discrepancies between the recovered and true mass in Fig.~\ref{fig:mass}.  In
prevision of future missions that will be able better control and reduce the
instrumental noise, we reanalyzed the same clusters using \chandra observations
for which the background accumulation time has been reduced by a factor of 100
implying that the effective background has been reduced by a factor of
100. Using these new data set we are able to extract cluster signal to larger
radii than before (we remind that we restricted our analysis to all the annuli
with at least 5000 net counts and with a flux from the source greater that
25\% of the total flux) that easily extend beyond $R_{500}$.  As before, we
find that the all the best fit values of the estimated X-ray mass
underestimate the true mass by a remarkable amount: by 20--40 per cent using
$M_{\beta}$ and by $\la 20$ per cent using $M_{HE}$ or the analytic functional
forms. Due to reduced statistical error, we can now see that these
discrepancies are in many cases statistically significant to more than
$3\sigma$ as clearly shown in Table~\ref{tab:mass2}.  It is worth saying that,
in contrast with Table~\ref{tab:mass} in this case the values at $R_{500}$ are
not extrapolation but measurements.

It is also worth noting the behaviour of the analytic masses in comparison
with the case of normal background reproduced (Table~\ref{tab:mass}): being
the errors largely reduced the true mass is always out of 3$\sigma$ and it
comes to be strongly underestimated. The masses recovered by the hydrostatic
equilibrium present the same characteristic, but for the clusters showing a
perturbed dynamical state because they are still conserving large error
bars. In particular, we investigated the reason of the understimation of
$M_{\rm HE}$ and we find that half of the total discrepancy is provided by
neglecting the kinetic energy still present as bulk motions of the
intra-cluster medium.  The remaining deviation arises from a systematically
lower measurement of the representative gas temperature.  In other words, in
these conditions of very low background, the deprojected spectral value is
systematically lower than the mass-weighted estimate from numerical
simulations by about 10 per cent, suggesting that a proper use of well-defined
gas temperature values in the hydrostatic equilibrium equation should allow a
more rigorous measurements of the total mass profile.

\section{SUMMARY AND DISCUSSION}  

We have investigated the bias on the X-ray mass estimates with mock long \chandra exposures of five galaxy clusters obtained from high-resolution
hydrodynamic simulations using an observational-like approach.  The five
objects have a spectroscopic-like temperature between $2.7$ and $11.4$ keV,
and are at redshift $z=0.175$ to fit their $R_{500}$ within \chandra
ACIS-S3 chip.  They are characterized by different dynamical states: two
objects are perturbed, one had a recent major merging, and two are fairly
relaxed.  These simulated clusters are processed with our X-ray Map Simulator,
\xmas, to produce realistic X-ray event files and images that are
analyzed with the goal of measuring the total mass profile.  To evaluate the
systematic effects present both in the assumptions of hydrostatic equilibrium
and of the $\beta$--model and in the observational techniques adopted in
recovering the mass measurement, we compare the estimated mass profiles to the
ones directly measured in the hydrodynamic simulations used as input for our
analysis.
 
The main results, shown in Fig.~\ref{fig:mass} and quoted in
Tables~\ref{tab:mass} and ~\ref{tab:mass2}, can be summarized as follows.
 
\begin{itemize} 
 
\item Due to the relatively high statistical errors mainly connected with the
high background level of the \chandra observations, the HE equation seems to
recover, within the errors, the true cluster mass profile of the simulated
clusters. This partially holds also for the analytic mass models NFW and RTM.
If however we reduce the background level by a factor of 100, we immediately
see, to a confidence level $>3\sigma$, that the reconstructed masses using
this technique are underestimated by 20\% (see Table~\ref{tab:mass2}).  In
this case, we find that the neglected kinetic pressure term can compensate for
about half of the observed deviations in relaxed systems, while the other half
is due to the underestimated measurement of the temperature in comparison to
the mass-weighted one.
 
\item The mass measurements reconstructed via the $\beta$--model are the worst
among the models considered in the present work since they show a systematic
underestimate, with typical deviations of about $40$ per cent at $R_{2500}$
and $R_{500}$.  $M_{\beta,\gamma}$ is 5-10 per cent lower than $M_{\beta}$ in
the outer regions ($r> R_{500}$) owing to the use of a too simplistic
polytropic equation state that relates temperature and density profiles.  The
use of the polytropic functional form of the temperature profile makes the
mass measurements less accurate than that provided from the isothermal
$\beta$--model.  The other sources of biases for the $\beta$--model mass come
from (i) its inaccurate description of the density profile, (ii) the poor
determination of the parameters describing the spatially-extended X-ray
emission.

\item A systematic effect, that influences the determination of the mass 
  through the $\beta$--model mass and the analytic formulae, is also produced 
  by the radial interval considered to constrain the value of $\beta$ of the 
  $\beta-$model and the concentration parameter in the NFW formula. We 
  stress that the observed $\beta$ and $c_{\rm NFW}$ will be always lower 
  and higher, respectively, than the values derived directly by fitting the 
  density of simulated clusters out to the virial radius \citep[see, e.g. the
  observational evidence in ][]{vikh.etal99,neumann05}. 
 
\item We conclude that the mass estimates based on the hydrostatic equilibrium
equation in combination with the temperature profile and those of the analytic
fits (NFW or RTM) provide a more robust mass estimate than the ones based on
the $\beta-$model.
 
\end{itemize} 
 
It is important to remarke that the data analysis performed in this paper has
been made under ideal conditions.  In fact we assumed to know precisely both
the background and the instrument response. Furthermore we used very long
exposures for all the simulated observations reaching a large number of net
counts.  It is clear that the uncertainties in the background and instrument
response which are inevitably present in real observations (as well as shorter
exposures) may easily make these discrepancies larger. Moreover we remind that
the physics adopted in the simulations biases our results on the temperature
and density profiles more significantly when the thermal structure of the
X-ray emitting plasma is more complex, in the sense that it becomes
problematic and ambiguous to represent the average properties of a gas which
presents a wide range in emission measures and temperatures.
 
Finally, we notice that these systematic effects on the X-ray total mass
propagate to the constraints on the cosmological parameters obtained by using
clusters as probes of the matter distribution in the Universe.  At the
present, statistical uncertainties associated to the paucity of available
high--redshift cluster samples are comparable to systematic uncertainties
arising from biases in cluster mass estimates \citep[see, e.g.,
][]{rosati2002}, as those discussed in this paper. As larger samples of
distant clusters will be available in the near future, it is clear that such
systematics will start to dominate over statistical errors.  In this respect,
extensive analyses, like the one presented here, will be of crucial relevance
to quantify possible biases in any procedure of cluster mass estimate.
However, the size of systematic biases and uncertainties, as calibrated from
hydrodynamic simulations, depends on the physical processes included in the
simulations themselves.  This highlights the fundamental role played by an
accurate numerical treatment of the complex physics of the ICM to calibrate
clusters from hydrodynamic simulations as precision tools for cosmology.
  
\section*{ACKNOWLEDGMENTS} 
 
The simulations have been realized with CPU time allocated at the ``Centro
Interuniversitario del Nord-Est per il Calcolo Elettronico'' (CINECA, Bologna)
thanks to grants from INAF and from the University of Trieste. This work has
been partially supported by the INFN grant PD-51. ER thanks for the
hospitality ESO and MPA in Garching, where part of the paper was written up.
PM acknowledge support from NASA grants GO4-5155X and GO5-6124X. We are
grateful to Giuseppe Murante, Volker Springel, Luca Tornatore, Paolo Tozzi,
Gus Evrard, Monique Arnaud, Hans Boehringer, Massimo Meneghetti and Rocco
Piffaretti for useful discussions.
 
\bibliographystyle{mn2e}

\begin{thebibliography}{}

\bibitem[\protect\citeauthoryear{{Allen}, {Schmidt} \& {Fabian}}{{Allen}
  et~al.}{2002}]{allen2002}
{Allen} S.~W.,  {Schmidt} R.~W.,    {Fabian} A.~C.,  2002, \mnras, 335, 256

\bibitem[\protect\citeauthoryear{{Anders} \& {Grevesse}}{{Anders} \&
  {Grevesse}}{1989}]{1989GeCoA..53..197A}
{Anders} E.,  {Grevesse} N.,  1989, \gca, 53, 197

\bibitem[\protect\citeauthoryear{{Arnaud}}{{Arnaud}}{1996}]{1996ASPC..101...17%
A}
{Arnaud} K.~A.,  1996, in ASP Conf. Ser. 101: Astronomical Data Analysis
  Software and Systems V {XSPEC: The First Ten Years}.
p.~17

\bibitem[\protect\citeauthoryear{{Balland} \& {Blanchard}}{{Balland} \&
  {Blanchard}}{1997}]{1997ApJ...487...33B}
{Balland} C.,  {Blanchard} A.,  1997, \apj, 487, 33

\bibitem[\protect\citeauthoryear{{Bartelmann} \& {Steinmetz}}{{Bartelmann} \&
  {Steinmetz}}{1996}]{bartelmann1996}
{Bartelmann} M.,  {Steinmetz} M.,  1996, \mnras, 283, 431

\bibitem[\protect\citeauthoryear{{Borgani} et~al.,}{{Borgani} et~al.}{2005}]{borgani.new}
{Borgani} S., et~al., 2005, preprint, astro-ph/0512506

\bibitem[\protect\citeauthoryear{{Borgani} et~al.,}{{Borgani}
  et~al.}{2004}]{borgani2004}
{Borgani} S.,  et~al., 2004, \mnras, 348, 1078

\bibitem[\protect\citeauthoryear{{Cavaliere} \& {Fusco-Femiano}}{{Cavaliere} \&
  {Fusco-Femiano}}{1976}]{cavaliere1976}
{Cavaliere} A.,  {Fusco-Femiano} R.,  1976, \aap, 49, 137

\bibitem[\protect\citeauthoryear{{Cowie}, {Henriksen} \& {Mushotzky}}{{Cowie}
  et~al.}{1987}]{1987ApJ...317..593C}
{Cowie} L.~L.,  {Henriksen} M.,    {Mushotzky} R.,  1987, \apj, 317, 593

\bibitem[\protect\citeauthoryear{{Dolag}, {Jubelgas}, {Springel}, {Borgani} \&
  {Rasia}}{{Dolag} et~al.}{2004}]{dolag.etal.04}
{Dolag} K.,  {Jubelgas} M.,  {Springel} V.,  {Borgani} S.,    {Rasia} E.,
  2004, \apjl, 606, L97

\bibitem[\protect\citeauthoryear{{Dolag}, {Vazza}, {Brunetti} \&
  {Tormen}}{{Dolag} et~al.}{2005}]{klaus05}
{Dolag} K.,  {Vazza} F.,  {Brunetti} G.,    {Tormen} G.,  2005, \mnras, 364,
  753

\bibitem[\protect\citeauthoryear{{Dorman}, {Arnaud} \& {Gordon}}{{Dorman}
  et~al.}{2003}]{2003xspec}
{Dorman} B.,  {Arnaud} K.~A.,    {Gordon} C.~A.,  2003, AAS/High Energy
  Astrophysics Division, 7,

\bibitem[\protect\citeauthoryear{{Dupke} \& {Bregman}}{{Dupke} \&
  {Bregman}}{2005}]{dupke2005}
{Dupke} R.~A.,  {Bregman} J.~N.,  2005, \apjs, 161, 224

\bibitem[\protect\citeauthoryear{{Eke}, {Cole} \& {Frenk}}{{Eke}
  et~al.}{1996}]{eke1996}
{Eke} V.~R.,  {Cole} S.,    {Frenk} C.~S.,  1996, \mnras, 282, 263

\bibitem[\protect\citeauthoryear{{Ettori}, {De Grandi} \& {Molendi}}{{Ettori}
  et~al.}{2002}]{ettori2002}
{Ettori} S.,  {De Grandi} S.,    {Molendi} S.,  2002, \aap, 391, 841

\bibitem[\protect\citeauthoryear{{Evrard}, {Metzler} \& {Navarro}}{{Evrard}
  et~al.}{1996}]{1996ApJ...469..494E}
{Evrard} A.~E.,  {Metzler} C.~A.,    {Navarro} J.~F.,  1996, \apj, 469, 494

\bibitem[\protect\citeauthoryear{{Gardini}, {Rasia}, {Mazzotta}, {Tormen}, {De
  Grandi} \& {Moscardini}}{{Gardini} et~al.}{2004}]{2004MNRAS.351..505G}
{Gardini} A.,  {Rasia} E.,  {Mazzotta} P.,  {Tormen} G.,  {De Grandi} S.,
  {Moscardini} L.,  2004, \mnras, 351, 505

\bibitem[\protect\citeauthoryear{{Henriksen} \& {Mushotzky}}{{Henriksen} \&
  {Mushotzky}}{1986}]{henriksen1986}
{Henriksen} M.~J.,  {Mushotzky} R.~F.,  1986, \apj, 302, 287

\bibitem[\protect\citeauthoryear{{Jubelgas}, {Springel} \& {Dolag}}{{Jubelgas}
  et~al.}{2004}]{2004MNRAS.351..423J}
{Jubelgas} M.,  {Springel} V.,    {Dolag} K.,  2004, \mnras, 351, 423

\bibitem[\protect\citeauthoryear{{Kay}, {Thomas}, {Jenkins} \& {Pearce}}{{Kay}
  et~al.}{2004}]{2004MNRAS.355.1091K}
{Kay} S.~T.,  {Thomas} P.~A.,  {Jenkins} A.,    {Pearce} F.~R.,  2004, \mnras,
  355, 1091

\bibitem[\protect\citeauthoryear{{Kriss}, {Cioffi} \& {Canizares}}{{Kriss}
  et~al.}{1983}]{1983ApJ...272..439K}
{Kriss} G.~A.,  {Cioffi} D.~F.,    {Canizares} C.~R.,  1983, \apj, 272, 439

\bibitem[\protect\citeauthoryear{{Mathiesen} \& {Evrard}}{{Mathiesen} \&
  {Evrard}}{2001}]{math_evr01}
{Mathiesen} B.~F.,  {Evrard} A.~E.,  2001, \apj, 546, 100

\bibitem[\protect\citeauthoryear{{Mazzotta}, {Rasia}, {Moscardini} \&
  {Tormen}}{{Mazzotta} et~al.}{2004}]{2004MNRAS.354...10M}
{Mazzotta} P.,  {Rasia} E.,  {Moscardini} L.,    {Tormen} G.,  2004, \mnras,
  354, 10

\bibitem[\protect\citeauthoryear{{Muanwong}, {Thomas}, {Kay} \&
  {Pearce}}{{Muanwong} et~al.}{2002}]{muanwong2002}
{Muanwong} O.,  {Thomas} P.~A.,  {Kay} S.~T.,    {Pearce} F.~R.,  2002, \mnras,
  336, 527

\bibitem[\protect\citeauthoryear{{Navarro}, {Frenk} \& {White}}{{Navarro}
  et~al.}{1995}]{NFW1995.1}
{Navarro} J.~F.,  {Frenk} C.~S.,    {White} S.~D.~M.,  1995, \mnras, 275, 720

\bibitem[\protect\citeauthoryear{{Navarro}, {Frenk} \& {White}}{{Navarro}
  et~al.}{1997}]{1997ApJ...490..493N}
{Navarro} J.~F.,  {Frenk} C.~S.,    {White} S.~D.~M.,  1997, \apj, 490, 493

\bibitem[\protect\citeauthoryear{Neumann}{Neumann}{2005}]{neumann05}
Neumann D.,  2005, \aap, 439, 465

\bibitem[\protect\citeauthoryear{{Norman} \& {Bryan}}{{Norman} \&
  {Bryan}}{1999}]{bn99}
{Norman} M.~L.,  {Bryan} G.~L.,  1999, LNP Vol.~530: The Radio Galaxy Messier
  87, 530, 106

\bibitem[\protect\citeauthoryear{{Rasia}, {Mazzotta}, {Borgani}, {Moscardini},
  {Dolag}, {Tormen}, {Diaferio} \& {Murante}}{{Rasia}
  et~al.}{2005}]{rasia.etal.05}
{Rasia} E.,  {Mazzotta} P.,  {Borgani} S.,  {Moscardini} L.,  {Dolag} K.,
  {Tormen} G.,  {Diaferio} A.,    {Murante} G.,  2005, \apjl, 618, L1

\bibitem[\protect\citeauthoryear{{Rasia}, {Tormen} \& {Moscardini}}{{Rasia}
  et~al.}{2004}]{2004MNRAS.351..237R}
{Rasia} E.,  {Tormen} G.,    {Moscardini} L.,  2004, \mnras, 351, 237

\bibitem[\protect\citeauthoryear{{Rosati}, {Borgani} \& {Norman}}{{Rosati}
  et~al.}{2002}]{rosati2002}
{Rosati} P.,  {Borgani} S.,    {Norman} C.,  2002, \araa, 40, 539

\bibitem[\protect\citeauthoryear{{Sarazin}}{{Sarazin}}{2002}]{2002mpgc.book...%
.1S}
{Sarazin} C.~L.,  2002, {The Physics of Cluster Mergers}.
ASSL Vol.~272: Merging Processes in Galaxy Clusters, pp 1--38

\bibitem[\protect\citeauthoryear{{Schindler}}{{Schindler}}{1996}]{1996A&A...30%
5..756S}
{Schindler} S.,  1996, \aap, 305, 756

\bibitem[\protect\citeauthoryear{{Springel}}{{Springel}}{2005}]{springel2005}
{Springel} V.,  2005, \mnras, 364, 1105

\bibitem[\protect\citeauthoryear{{Springel} \& {Hernquist}}{{Springel} \&
  {Hernquist}}{2002}]{2002MNRAS.333..649S}
{Springel} V.,  {Hernquist} L.,  2002, \mnras, 333, 649

\bibitem[\protect\citeauthoryear{{Springel} \& {Hernquist}}{{Springel} \&
  {Hernquist}}{2003}]{springel2003}
{Springel} V.,  {Hernquist} L.,  2003, \mnras, 339, 289

\bibitem[\protect\citeauthoryear{Springel, Yoshida \& White}{Springel
  et~al.}{2001}]{SP01.1}
Springel V.,  Yoshida N.,    White S.,  2001, New Astronomy, 6, 79

\bibitem[\protect\citeauthoryear{Tormen, Bouchet \& White}{Tormen
  et~al.}{1997}]{1997MNRAS.286..865T}
Tormen G.,  Bouchet F.,    White S.,  1997, MNRAS, 286, 865

\bibitem[\protect\citeauthoryear{{Vazza}, {Tormen}, {Cassano}, {Brunetti} \&
  {Dolag}}{{Vazza} et~al.}{2006}]{fra06}
{Vazza} F.,  {Tormen} G.,  {Cassano} R.,  {Brunetti} G.,    {Dolag} K.,  2006,
  ArXiv Astrophysics e-prints

\bibitem[\protect\citeauthoryear{{Vikhlinin}}{{Vikhlinin}}{2005}]{vikhlinin200%
5}
{Vikhlinin} A.,  2005, preprint, astro-ph/0504098

\bibitem[\protect\citeauthoryear{{Vikhlinin} et~al.,}{{Vikhlinin}
  et~al.}{2005}]{vikh05mass}
{Vikhlinin} A.,  et~al., 2005, preprint, astro-ph/0507092

\bibitem[\protect\citeauthoryear{{Vikhlinin}, {Forman} \& {Jones}}{{Vikhlinin}
  et~al.}{1999}]{vikh.etal99}
{Vikhlinin} A.,  {Forman} W.,    {Jones} C.,  1999, \apj, 525, 47

\bibitem[\protect\citeauthoryear{{Voit}}{{Voit}}{2005}]{voit2005}
{Voit} G.~M.,  2005, Reviews of Modern Physics, 77, 207

\bibitem[\protect\citeauthoryear{{Yoshida}, {Colberg}, {White}, {Evrard},
  {MacFarland}, {Couchman}, {Jenkins}, {Frenk}, {Pearce}, {Efstathiou},
  {Peacock} \& {Thomas}}{{Yoshida} et~al.}{2001}]{yoshida.etal.01}
{Yoshida} N.,  {Colberg} J.,  {White} S.~D.~M.,  {Evrard} A.~E.,  {MacFarland}
  T.~J.,  {Couchman} H.~M.~P.,  {Jenkins} A.,  {Frenk} C.~S.,  {Pearce} F.~R.,
  {Efstathiou} G.,  {Peacock} J.~A.,    {Thomas} P.~A.,  2001, \mnras, 325, 803

\bibitem[\protect\citeauthoryear{{Zhang}, {B{\" o}hringer}, {Mellier},
  {Soucail} \& {Forman}}{{Zhang} et~al.}{2005}]{zhang2005}
{Zhang} Y.-Y.,  {B{\" o}hringer} H.,  {Mellier} Y.,  {Soucail} G.,    {Forman}
  W.,  2005, \aap, 429, 85

\bibitem[\protect\citeauthoryear{{Zwicky}}{{Zwicky}}{1933}]{zwicky1933}
{Zwicky} F.,  1933, Helvetica Physica Acta, 6, 110

\end{thebibliography}

\end{document}